\documentclass[12pt]{article}

\newcommand{\ri}{\rightarrow}

\newcommand{\De}{\Delta}

\title{Toward quantum-like modeling of financial processes}

\author{Olga Choustova\\International Center for Mathematical
Modeling \\ in Physics and Cognitive Sciences,\\
University of V\"axj\"o, S-35195, Sweden\\
Olga.Choustova@vxu.se}

\begin{document}
\maketitle

\begin{abstract}
We apply methods of quantum mechanics for mathematical modeling of
price dynamics at the financial market.  We propose to describe
behavioral financial factors (e.g., expectations of traders) by
using the pilot wave (Bohmian) model of quantum mechanics.
Trajectories of prices are determined by two financial potentials:
classical-like $V(q)$ ("hard" market conditions, e.g., natural
resources) and quantum-like $U(q)$ (behavioral market conditions).
On the one hand, our Bohmian model is a quantum-like model for the
financial market, cf. with works of W. Segal, I. E. Segal, E. Haven,
E. W. Piotrowski, J. Sladkowski. On the other hand, (since Bohmian
mechanics provides the possibility to describe individual price
trajectories) it belongs to the domain of extended research on
deterministic dynamics for financial assets (C.W. J. Granger, W.A.
Barnett, A. J. Benhabib, W.A. Brock, C. Sayers, J. Y. Campbell, A.
W. Lo, A. C. MacKinlay, A. Serletis, S. Kuchta, M. Frank, R. Gencay,
T. Stengos, M. J. Hinich, D. Patterson, D. A. Hsieh, D. T. Caplan,
J.A. Scheinkman, B. LeBaron and many others).
\end{abstract}

\section{Deterministic and Stochastic Models for Financial Market}

\subsection{Stochastic models and the efficient market hypothesis}

In economics and financial theory, analysts use random walk and more
general martingale techniques to model behavior of asset prices, in
particular share prices on stock markets, currency exchange rates
and commodity prices. This practice has its basis in the presumption
that investors act {\it rationally and without bias,} and that at
any moment they estimate the value of an asset based on future
expectations. Under these conditions, all existing information
affects the price, which changes only when new information comes
out. By definition, {\bf new information appears randomly and
influences the asset price randomly.} Corresponding continuous time
models are based on stochastic processes (this approach was
initiated in the thesis of L. Bachelier \cite{BA} in 1890), see,
e.g., the books of  R. N. Mantegna and H. E. Stanley \cite{MAN} and
A. Shiryaev \cite{Sh} for historical and mathematical details.

This practice was formalized through the {\it efficient market
hypothesis} which  was formulated in the sixties, see P. A. Samuelson
\cite{SM} and E. F. Fama \cite{Fam} for details:

\medskip

{\it A market is said to be efficient in the determination of the
most rational price if all the available information is instantly
processed when it reaches the market and it is immediately reflected
in a new value of prices of the assets traded.}

\medskip

Mathematically the efficient market hypothesis was supported by
investigations of Samuelson \cite{SM}. Using the hypothesis of
rational behavior and market efficiency he was able to demonstrate
how $q_{t+1},$  the expected value of price of a given asset at time
$t+1,$ is related to the previous values of prices $q_0, q_1,...,
q_t$ through the relation
\begin{equation}
\label{GT} E (q_{t+1}\vert q_0, q_1,..., q_t)= q_t.
\end{equation}
Typically there is introduced the $\sigma$-algebra ${\cal F}_t$
generated by random variables $q_0, q_1,..., q_t.$ The condition
(\ref{GT}) is written in the form: \begin{equation} \label{SMZ} E
(q_{t+1}\vert {\cal F}_t)= q_t.
\end{equation}
Stochastic processes of such a type are called martingales
\cite{Sh}. Alternatively, the martingale model for the financial
market implies that the $$(q_{t+1}-q_{t})$$ is a ``fair game'' (a game
which is neither in your favor nor your opponent's):
\begin{equation}
\label{SMZR} E (q_{t+1}- q_{t} \vert {\cal F}_t)= 0.
\end{equation}
On the basis of information, ${\cal F}_t,$ which is available at the
moment $t,$ one cannot expect neither $$E (q_{t+1}- q_{t} \vert {\cal
F}_t)>0$$ nor $$E (q_{t+1}- q_{t} \vert {\cal F}_t)<0.$$

\subsection{Deterministic models for dynamics of prices}

First we remark that empirical studies have demonstrated that prices
do not completely follow random walk. Low serial correlations
(around 0.05) exist in the short term; and slightly stronger
correlations over the longer term. Their sign and the strength
depend on a variety of factors, but transaction costs and bid-ask
spreads generally make it impossible to earn excess returns.
Interestingly, researchers have found that some of the biggest
prices deviations from random walk result from seasonal and temporal
patterns, see the book \cite{MAN}.

There are also a variety of arguments, both theoretical and obtained
on the basis of statistical analysis of data, which question the
general martingale model (and hence the efficient market
hypothesis), see, e.g., \cite{Bar}--\cite{Hs}.  It is important to
note that efficient markets imply there are no exploitable profit
opportunities. If this is true then trading on the stock market is a
game of chance and not of any skill, but traders buy assets they
think are unevaluated at the hope of selling them at their true
price for a profit. If market prices already reflect all information
available, then where does the trader draw this privileged
information from? Since there are thousands of very well informed,
well educated asset traders, backed by many data researchers, buying
and selling securities quickly, logically assets markets should be
very efficient and profit opportunities should be minimal. On the
other hand, we see that there are many traders whom successfully use
their opportunities and perform continuously very successful
financial operations, see the book of G. Soros \cite{S} for
discussions.\footnote{It seems that G.Soros is sure he does not work
at efficient markets.} There were also performed intensive
investigations on testing that the real financial data can be really
described by the martingale model, see \cite{Bar}--\cite{Hs}.
Roughly speaking people try to understand on the basis of available
financial data:

\medskip

Do financial asset returns behave randomly (and hence they are
unpredictable) or deterministically (and in the latter case one may
hope to predict them and even to construct a deterministic dynamical
system which would at least mimic dynamics of the financial market)?

\medskip

Predictability of financial asset returns is a broad and very active
research topic and a complete survey of the vast literature is
beyond the scope of this work. We shall note, however, that there is
a rather general opinion that financial asset returns are
predictable, see \cite{Bar}--\cite{Hs}.

\subsection{Behavioral finance and behavioral economics}

We point out that there is no general consensus on the validity of
the efficient market hypothesis. As it was pointed out in
\cite{Cam}: ``... econometric advances and empirical evidence seem to
suggest that financial asset returns are predictable to some degree.
Thirty years ago this would have been tantamount to an outright
rejection of market efficiency. However, modern financial economics
teaches us that others, perfectly rational factors may account for
such predictability. The fine structure of securities markets and
frictions in trading process can generate predictability.
Time-varying expected returns due to changing business conditions
can generate predictability. A certain degree of predictability may
be necessary to reward investors for bearing certain dynamic
risks.''

Therefore it would be natural to develop approaches which are not based on the
assumption that investors act {\it rationally and without bias} and
that, consequently, new information appears randomly and influences
the asset price randomly. In particular, there are two well
established (and closely related ) fields of research  {\it
behavioral finance and behavioral economics}  which apply scientific
research on human and social cognitive and emotional
biases\footnote{Cognitive bias is any of a wide range of observer
effects identified in cognitive science, including very basic
statistical and memory errors that are common to all human beings
and drastically skew the reliability of anecdotal and legal
evidence. They also significantly affect the scientific method which
is deliberately designed to minimize such bias from any one
observer. They were first identified by Amos Tversky and Daniel
Kahneman as a foundation of behavioral economics, see, e.g., \cite{TR}. Bias arises from
various life, loyalty and local risk and attention concerns that are
difficult to separate or codify. Tversky and Kahneman claim that
they are at least partially the result of problem-solving using
heuristics, including the availability heuristic and the
representativeness.} to better understand economic decisions and how
they affect market prices, returns and the allocation of resources.
The fields are primarily concerned with the rationality, or lack
thereof, of economic agents. Behavioral models typically integrate
insights from psychology with neo-classical economic theory.
Behavioral analysis are mostly concerned with the effects of market
decisions, but also those of public choice, another source of
economic decisions with some similar biases.

Since the 1970s, the intensive exchange of information in the world
of finances has become one of the main sources determining dynamics
of prices. Electronic trading (that became the most  important part
of the environment of the major stock exchanges) induces huge
information flows between traders (including foreign exchange
market). Financial contracts are performed at a new time scale that
differs essentially from the old "hard" time scale that was
determined by the development of the economic basis of the financial
market. Prices at which traders are willing to buy (bid quotes) or
sell (ask quotes) a financial asset are not only determined by the
continuous development of industry, trade, services, situation at
the market of natural resources and so on. Information (mental,
market-psychological) factors play a very important (and in some
situations crucial) role in price dynamics. Traders performing
financial operations work as a huge collective cognitive system.
Roughly speaking classical-like dynamics of prices (determined) by
"hard" economic factors are permanently perturbed by additional
financial forces, mental (or market-psychological) forces, see the
book of J. Soros \cite{S}.

\subsection{Quantum-like model for behavioral finance}

In this thesis we develop a new approach that is not based on the
assumption that investors act rationally and without bias and that,
consequently, new information appears randomly and influences the
asset price randomly. Our approach can be considered as a special
econophysical \cite{MAN} model in the domain of  behavioral finance.
In our approach information about the financial market (including
expectations of agents of the financial market) is described by an
{\it information field} $\psi(q)$ -- {\it financial wave.} This
field evolves deterministically 
perturbing the dynamics of prices of stocks and options. 
Dynamics is given by
Schr\"odinger's equation on the space of prices of shares.
Since
the psychology of agents of the financial market gives an important
contribution into the financial wave $\psi(q),$ our model can be
considered as a special {\it psycho-financial model.}

This thesis can be also considered as a contribution into
applications of quantum mechanics outside microworld, see \cite{AR},
\cite{AC}, \cite{KHR}. This thesis is fundamentally based on
investigations of D. Bohm, B. Hiley, and  P. Pylkk\"anen
\cite{BOHM1}, \cite{HILEY} on the {\it active information}
interpretation of Bohmian mechanics \cite{BOHM}, \cite{HOL} and its
applications to cognitive sciences, see also Khrennikov \cite{KHR}.

In this thesis we use methods of Bohmian mechanics to simulate
dynamics of prices  at the financial market. We start with the
development of the classical Hamiltonian formalism on the
price/price-change phase space  to describe the classical-like
evolution of prices. This classical dynamics of prices is determined
by "hard" financial conditions (natural resources, industrial
production, services and so on). These conditions as well as "hard"
relations between traders at the financial market are mathematically
described by the classical financial potential. As we have already
remarked, at the real financial market "hard" conditions are not the
only source  of price changes. The information and market psychology
play important (and sometimes determining) role in price dynamics.

We propose to describe those"soft" financial factors by using the
pilot wave (Bohmian) model of quantum mechanics. The theory of
financial mental (or psychological) waves is used to take into
account market psychology. The real trajectories of prices are
determined (by the financial analogue of the second Newton law) by
two financial potentials: classical-like ("hard" market conditions)
and quantum-like ("soft" market conditions).

Our quantum-like model of financial processes was strongly motivated
by consideration by J. Soros \cite{S} of the financial market as a
complex cognitive system. Such an approach he called the theory of
{\it reflexivity.} In this theory there is  a large difference
between market that is "ruled" by only "hard" economical factors and
a market where mental factors play the crucial role (even changing
the evolution of the "hard" basis, see \cite{S}).

J. Soros rightly remarked that the "non mental" market evolves due to
classical random fluctuations. However, such fluctuations do not
provide an adequate description of mental market. He proposed to use
an analogy with quantum theory. However, it was noticed that directly
quantum formalism could not be applied to the financial market
\cite{S}. Traders differ essentially from elementary particles.
Elementary particles behave stochastically due to perturbation
effects provided by measurement devices, cf. \cite{H}, \cite{D}.

According to J. Soros, traders at the financial market behave
stochastically due to free will of individuals. Combinations of a
huge number of free wills of traders produce additional
stochasticity at the financial market that could not be reduced to
classical random fluctuations (determined by non mental factors).
Here J. Soros followed to the conventional (Heisenberg, Bohr, Dirac,
see, e.g.,  \cite{H}, \cite{D}) viewpoint to the origin of quantum
stochasticity. However, in the Bohmian approach (that is
nonconventional one) quantum statistics is induced by the action of
an additional potential, quantum potential, that changes classical
trajectories of elementary particles. Such an approach gives the
possibility to apply quantum formalism to the financial market.

We remark that applications of the pilot-wave theory to financial option pricing
were considered by E. Haven in \cite{HAA2}, see also \cite{HA}.

\subsection{Review on investigations in quantum econophysics}

There were performed numerous investigations on applying quantum
methods to financial market, see, e.g., E. Haven \cite{HA1}--
\cite{HA3}, that were not directly coupled to behavioral modeling,
but based on the general concept that randomness of the financial
market can be better described by the quantum mechanics, see, e.g., W. Segal and I. E. Segal
\cite{SS}: "A natural explanation for extreme irregularities in the
evolution of prices in financial markets is provided by quantum
effects."

Non-Bohmian quantum models for the financial market (in particular, based on quantum games) 
were developed by
E. W. Piotrowski, J. Sladkowski, M. Schroeder, A. Zambrzycka, see 
\cite{PPP}--\cite{PPP2}. Some of those models can be also
considered as behavioral quantum-like models.

An interesting contribution to behavioral quantum-like modeling is 
theory of non-classical measurements
in behavioral sciences (with applications to economics) which was
developed by  V. I. Danilov and A. Lambert-Mogiliansky  \cite{DAN}, \cite{DAN1}.

\section{A Brief Introduction to Bohmian Mechanics}

In this section we present the basic notions of Bohmian mechanics.
This is a special model of quantum mechanics in that, in the
opposition to the conventional Copenhagen interpretation, quantum
particles (e.g., electrons) have  well defined trajectories in
physical space.

By the conventional Copenhagen interpretation (that  was created by
N. Bohr and W. Heisenberg) quantum particles do not have
trajectories in physical space. Bohr and Heisenberg motivated such a
viewpoint to quantum physical reality by the Heisenberg uncertainty
relation:

\begin{equation}
\label{HW} \Delta q \Delta p \geq h/2
\end{equation}
where $h$ is the Planck constant, $q$ and $p$ are the position and
momentum, respectively, and $\Delta q$ and $\Delta p$ are
uncertainties in determination of $q$ and $p.$ Nevertheless, David
Bohm demonstrated \cite{BOHM}, see also \cite{HOL}, that, in spite
of Heisenberg's uncertainty relation (\ref{HW}), it is possible to
construct a quantum model in that trajectories of quantum particles
are well defined. Since this thesis is devoted to mathematical models
in economy and not to physics, we would not go deeper into details.
We just mention that the root of the problem lies in different
interpretations of Heisenberg's uncertainty relation (\ref{HW}). If
one interpret $\Delta q$ and $\Delta p$ as uncertainties for the
position and momentum of an individual quantum particle (e.g., one
concrete electron) then (\ref{HW}), of course implies that it is
impossible to create a model in that the trajectory $q(t), p(t)$ is
well defined. On the other hand, if one interpret $\Delta q$ and
$\Delta p$ as {\it statistical deviations}
\begin{equation}
\label{SD} \Delta q=\sqrt{E(q-Eq)^2},\; \;  \Delta
p=\sqrt{E(p-Ep)^2},
\end{equation}
then there is no direct contradiction between Heisenberg's
uncertainty  relation (\ref{HW}) and the possibility to consider
trajectories. There is a place to such models as Bohmian mechanics.
Finally, we remark (but without comments) that in real experiments
with quantum systems, one always uses the statistical
interpretation (\ref{SD}) of $\Delta q$ and $\Delta p$.

We emphasize that the conventional quantum formalism cannot say
anything about the individual quantum particle. This formalism
provides only statistical predictions on huge ensembles of
particles. Thus Bohmian mechanics provides a better description of
quantum reality, since there is the possibility to describe
trajectories of individual particles. However, this great advantage
of Bohmian mechanics was not explored so much in physics. Up to now
there have not been done experiments that would distinguish
predictions of Bohmian mechanics and conventional quantum mechanics.

In this thesis we shall show that the mentioned advantages  of
Bohmian mechanics can be explored in applications to the financial
market. In the latter case it is really possible to observe the
trajectory of the price or price-change dynamics. Such a trajectory
is described by equations of the mathematical formalism of Bohmian
mechanics.

We now present the detailed derivation of the equations of motion of
a quantum particle in the Bohmian model of quantum mechanics.
Typically in physical books it is presented very briefly. But, since
this thesis is oriented to  economists and mathematicians, who are
not so much aware about quantum physics, we shall present all
calculations. The dynamics of the wave function $\psi (t, q)$ is
described by Schr\"odinger's equation
\begin{equation}
\label{SE1} i \; h \frac{\partial \psi}{\partial t} (t, q)= -
\frac{h^2}{2m} \frac{\partial^2 \psi}{\partial q^2} (t, q) + V(q)
\psi(t, q)
\end{equation}
Here $\psi(t, q)$ is a complex valued function. At the moment we
prefer not to discuss  the conventional probabilistic interpretation of
$\psi(t, q).$  We consider $\psi (t, q)$ as just a
field.\footnote{We recall that by the probability interpretation of
$\psi(t, q)$ (which was proposed by Max Born) the quantity $\vert
\psi(t, q)\vert^2$ gives the probability to find a quantum particle
at the point $q$ at the moment $t.$}

We consider the one-dimensional case, but the generalization to the
multidimensional case, $q=(q_1,  \ldots, q_n),$ is straightforward.
Let us write the wave function $\psi(t, q)$ in the  following form:
\begin{equation}
\label{ZW} \psi(t, q)=R(t, q) e^{i \frac{S (t, q)}{h}}
\end{equation}
where $R(t, q)= \vert \psi(t, q) \vert$ and $\theta(t, q)=S(t, q)/h$
is the argument of the complex number $\psi(t, q).$

We put (\ref{ZW}) into Schr\"odinger's equation (\ref{SE1}). We have
$$i h \frac{\partial \psi}{\partial t}=ih \Big(\frac{\partial R}{\partial t} e^{\frac{i S}{h}}
+ \frac{i R}{h}\frac{\partial S}{\partial t} e^{\frac{i S}{h}}\Big)
= i h \frac{\partial R}{\partial t} e^{\frac{i S}{h}} - R
\frac{\partial S}{\partial t} e^{\frac{i S}{h}}$$

and

$$\frac{\partial \psi}{\partial q}=\frac{\partial R}{\partial q} e^{\frac{i S}{h}} +
\frac{i R}{h} \frac{\partial S}{\partial q} e^{\frac{i S}{h}}$$

and hence:
$$\frac{\partial^2 \psi}{\partial q^2}=
\frac{\partial^2 R}{\partial q^2} e^{\frac{i S}{h}} +
\frac{2i}{h}\frac{\partial R}{\partial q} \frac{\partial S}{\partial
q} e^{\frac{i S}{h}} + \frac{i R}{h} \frac{\partial^2 S}{\partial
q^2} e^{\frac{iS}{h}} - \frac{R}{h^2} \Big(\frac{\partial
S}{\partial q}\Big)^2 e^{\frac{iS}{h}}$$

We obtain the differential equations:

\begin{equation}
\label{B1} \frac{\partial R}{\partial t}=\frac{-1}{2m} \Big(2
\frac{\partial R}{\partial q} \frac{\partial S}{\partial q} + R
\frac{\partial^2 S}{\partial q^2}\Big),
\end{equation}

\begin{equation}
\label{S2} -R \frac{\partial S}{\partial t} = -\frac{h^2}{2m}
\Big(\frac{\partial^2 R}{\partial q^2} - \frac{R}{h^2}
\Big(\frac{\partial S}{\partial q}\Big)^2\Big) + VR.
\end{equation}

By multiplying the  right and left-hand sides of the equation
(\ref{B1}) by $2R$ and using the trivial equalities: $$\frac{\partial R^2}{\partial t}= 2R
\frac{\partial R}{\partial t}$$ and $$\frac{\partial}{\partial q} (R^2
\frac{\partial S}{\partial q})= 2 R \frac{\partial R}{\partial q}
\frac{\partial S}{\partial q} + R^2 \frac{\partial^2 S}{\partial
q^2},$$
we derive the equation for $R^2$:

\begin{equation}
\label{S3} \frac{\partial R^2}{\partial t} + \frac{1}{m}
\frac{\partial}{\partial q} (R^2 \frac{\partial S}{\partial q})=0.
\end{equation}

We remark that if one uses the Born's probabilistic interpretation
of the wave function, then $$R^2(t,x)=|\psi(t,x)|^2$$ gives the probability.
Thus the equation (\ref{S3}) is the equation describing the dynamics
of the probability distribution (in physics it is called the
continuity equation).

The second equation can be written in the form:

\begin {equation}
\label{S4} \frac{\partial S}{\partial t} + \frac{1}{2m}
\Big(\frac{\partial S}{\partial q}\Big)^2 + \Big(V-\frac{h^2}{2 m R}
\frac{\partial^2 R}{\partial q^2}\Big)=0.
\end{equation}

Suppose that $$\frac{h^2}{2m} < < 1$$ and that the contribution of
the term $$\frac{h^2}{2 m R} \frac{\partial^2 R}{\partial q^2}$$ can
be neglected. Then we obtain the equation:

\begin{equation}
\label{S5} \frac{\partial S}{\partial t} + \frac{1}{2m}
\Big(\frac{\partial S}{\partial q}\Big)^2 + V=0.
\end{equation}

From the classical mechanics,  we know that this is the classical
Hamilton-Jacobi equation which corresponds to the dynamics of
particles:

\begin{equation}
\label{SG} p= \frac{\partial S}{\partial q} \;{\rm {or }} \;m \dot
q=\frac{\partial S}{\partial q},
\end{equation}

where particles moves normal to the surface $S=const.$

David Bohm proposed to interpret the equation (\ref{S4}) in the same
way. But we see that in this equation the classical potential $V$ is
perturbed by an additional "quantum potential" $$U =\frac{h^2}{2 m
R} \frac{\partial^2 R}{\partial q^2}.$$

Thus in the Bohmian mechanics the motion of a particle is described
by the usual Newton equation, but with the force corresponding to
the combination of the classical potential $V$ and the quantum one
$U:$

\begin{equation}
\label{QP} m \frac{dv}{dt}=-(\frac{\partial V}{\partial
q}-\frac{\partial U}{\partial q})
\end{equation}

The crucial point is that the potential $U$ is by itself driven by a
field equation - Schr\"odinger's equation (\ref{SE1}). Thus the
equation (\ref{QP}) can not be considered as just the Newton
classical dynamics (because the potential $U$ depends on $\psi$ as a
field parameter). We shall call (\ref{QP}) the {\it Bohm-Newton
equation.}

We remark that typically in books on Bohmian mechanics \cite{BOHM},
\cite{HOL} it is emphasized that the equation (\ref{QP}) is nothing
else than the ordinary Newton equation. This make impression that
the Bohmian approach give the possibility to reduce quantum
mechanics to ordinary classical mechanics. However, this is not the
case. The equation (\ref{QP}) does not provide the complete
description of dynamics of a systems. Since, as was pointed out, the
quantum potential $U$ is determined through the wave function $\psi$
and the latter evolves according to the Schr\"odinger equation, the
dynamics given by Bohm-Newton equation can not be considered
independent of the Schr\"odinger's dynamics.

\section{Classical Econophysical Model for Financial Market}

\subsection{Financial phase-space}

Let us consider a mathematical model in that a huge number of agents
of the financial market interact with one another and take into
account  external economic (as well as political, social and even
meteorological) conditions in order to determine the  price to buy
or sell financial assets. We consider the trade with shares of some
corporations (e.g., VOLVO, SAAB, IKEA,...).\footnote{Similar models
can be developed for trade with options, see E. Haven \cite{HA} for
the Bohmian financial wave model for portfolio.}

We consider a {\it{price system of coordinates.}} We enumerate
corporations which did emissions of shares at the financial market
under consideration: $j=1,2,....,n$ (e.g., VOLVO:$j=1,$ SAAB:$j=2,$
IKEA:$j=3$,...). There can be introduced the $n$-dimensional
configuration space $Q=R^n$ of prices, $$q=(q_1, \ldots, q_n),$$
where $q_j$ is the price of a share of the $j$th corporation. Here
$R$ is the real line. Dynamics of prices is described by the
trajectory $$q(t)=(q_1(t), \ldots, q_n(t))$$ in the configuration
price space $Q.$

Another variable under the consideration is the {\it price change
variable}: $$v_j(t)=\dot q_j(t)=\lim_{\De t\ri 0}\frac{q_j(t + \De
t)-q_j(t)}{\De t},$$ see, for example, the book \cite{MAN} on the
role of the price change description. In real models we consider the
discrete time scale $$\De t, 2\De t, \ldots.$$ Here we should use
a discrete price change variable $$\delta q_j(t)=q_j(t+\De
t)-q_j(t).$$

We denote the space of price changes  (price velocities)  by the symbol $$V(\equiv R^n)$$
with coordinates $$v=(v_1, \ldots, v_n).$$ As in classical physics, it is useful to
introduce the phase space $Q \times V = R^{2n},$ namely the {\it
price phase space.} A pair 

$(q, v)$ = (price, price change) 

is called the {\it state of the financial market. }

Later we shall consider
quantum-like states of the financial market. A state $(q,v)$ which we consider at the moment  is a
classical state.

We now introduce an analogue $m$ of mass as the number of items (in
our case shares) that a trader emitted to the market.\footnote{
`Number' is  a natural number $m=0,1, \ldots,$ -- the price of
share, e.g., in the US-dollars. However, in a mathematical model it
can be convenient to consider real $m.$ This can be useful for
transitions from one currency to another.} We call $m$ the {\it
financial mass.} Thus each trader $j$ (e.g., VOLVO) has its own
financial mass $m_j$ (the size of the emission of its shares). The
total price of the emission performed by the $j$th trader  is equal
to $T_j = m_j q_j$ (this is nothing else than {\it market
capitalization}). Of course, it depends on time: $T_j(t) = m_j
q_j(t).$ To simplify considerations, we consider a market at that
any emission of shares is of the fixed size, so $m_j$ does not
depend on time. In principle, our model can be generalized to
describe a market with time-dependent financial masses,
$m_j=m_j(t).$

We also introduce {\it{financial energy}} of the market as a
function $$H: Q \times V\ri R.$$ If we use the analogue with classical
mechanics. (Why not? In principle, there is not so much difference
between motions in "physical space" and "price space".), then we
could consider (at least for mathematical modeling) the financial
energy of the form:
\begin{equation}
\label{E} H(q,v)=\frac{1}{2}\sum_{j=1}^{n}m_j v_j^2 + V(q_1, \ldots,
q_n).
\end{equation}
Here $$K(q,v)=\frac{1}{2}\sum_{j=1}^{n}m_j v_j^2$$ 
is the {\it kinetic
financial energy} and $$V(q_1, \ldots, q_n)$$ is the potential
financial energy, $m_j$ is the financial mass of $j$th trader.

The kinetic financial energy represents efforts of agents of
financial market to change prices: higher price changes induce
higher kinetic financial energies. If the corporation $j_1$ has
higher financial mass than the corporation $j_2,$ so $m_{j_1} >
m_{j_2},$ then the same change of price, i.e., the same financial
velocity $v_{j_1}= v_{j_2},$ is characterized by higher kinetic
financial energy: $K_{j_1} > K_{j_2}.$ We also remark that high
kinetic financial energy characterizes rapid changes of the
financial situation at market. However, the kinetic financial energy
does not give the attitude of these changes. It could be rapid
economic growth as well as recession.

The {\it potential financial energy} $V$ describes the interactions
between traders $j=1,....,n$ (e.g., competition between NOKIA and
ERICSSON) as well as external economic conditions (e.g., the price
of oil and gas) and even meteorological conditions (e.g., the
weather conditions in Louisiana and Florida). For example, we can
consider the simplest interaction potential: $$V(q_1, \ldots,
q_n)=\sum_{j=1}^{n}(q_i-q_j)^2.$$ The difference $|q_1-q_j|$ between
prices is the most important condition for {\it arbitrage.}

We could never  take into account all economic and other conditions
that  have influences to the market. Therefore by using some
concrete potential $V(q)$ we consider the very idealized model of
financial processes. However, such an approach is standard for
physical modeling where we also consider idealized mathematical
models of real physical processes.

\subsection{Classical dynamics}

 We apply the Hamiltonian dynamics on
the price phase space. As in classical mechanics for material
objects, we introduce a new variable $$p=m v,$$ the {\it price
momentum} variable. Instead of the price change vector $$v=(v_1,
\ldots, v_n),$$ we  consider the price momentum vector $$p=(p_1,
\ldots, p_n), p_j=m_j v_j.$$ The space of price momentums is denoted
by the symbol $P.$ The space $$\Omega=Q\times P$$ will be also called
the price phase space. {\it Hamiltonian equations} of motion on the
price phase space have the form: $$ \dot q=\frac{\partial
H}{\partial p_j}, \dot p_j=-\frac{\partial H}{\partial q_j}, j=1,
\ldots, n.$$

If the financial energy has form ({\ref{E}}) then the Hamiltonian
equations have the form $$\dot q_j=\frac{p_j}{m_j}=v_j, \dot
p_j=-\frac{\partial V}{\partial q_j}.$$ The latter equation can be
written in the form: $$m_j \dot{v}_j=- \frac{\partial V}{\partial
q_j}.$$ The quantity $$\dot{v}_j=\lim_{\Delta t\ri 0}
\frac{v_j(t+\Delta t)-v_j(t)}{\Delta t}$$ is natural to call the
{\it price acceleration} (change of price change). The quantity
$$f_j(q)=-\frac{\partial V}{\partial q_j}$$ is called the (potential)
financial force. We get the financial variant of the second Newton
law:
\begin{equation}
\label{N} m\dot{v}= f
\end{equation}

\medskip

{\bf "The  product of the financial mass and the  price acceleration
is equal to the financial force."}

\medskip

In fact, the Hamiltonian evolution is determined by the following
fundamental property of the financial energy: {\it The financial
energy is not changed in the process of Hamiltonian evolution:}
$$H(q_1(t), \ldots, q_n(t), p_1(t), \ldots, p_n(t)=H(q_1(0), \ldots
q_n(0), p_1(0), \ldots, p_n(0)).$$ 

We need not restrict our
considerations to financial energies of form ({\ref{E}}). First of
all external (e.g. economic) conditions as well as the character of
interactions between traders at the market depend strongly on time.
This must be taken into account by considering time dependent
potentials: $$V=V(t, q).$$ 

Moreover, the assumption that the financial
potential depends only on prices, $V=V(t, q),$ is not so natural for
the modern financial market. Financial agents  have the complete
information on price changes. This information is  taken into
account by traders for acts of arbitrage, see \cite{MAN} for the
details. Therefore, it can be useful to consider potentials that
depend not only on prices, but also on price changes: $$V = V(t, q,
v)$$ or in the Hamiltonian framework: $$V=V(t, q, p).$$ In such a case
the financial force is not potential. Therefore, it is also useful
to consider the financial second Newton law for general financial
forces: $$m \dot {v}= f(t, q, p).$$

\medskip

{\bf Remark 1.} (On the form of the kinetic financial energy) We
copied the form of kinetic energy from classical mechanics for
material objects. It may be that such a form of kinetic financial
energy is not justified by real financial market. It might be better
to consider our choice of the kinetic financial energy as just the
basis for mathematical modeling (and looking for other
possibilities).

\medskip

{\bf Remark 2.} (Domain of price-dynamics) It is natural to consider
a model in that all prices are nonnegative, $q_j(t)\geq 0.$
Therefore financial Hamiltonian dynamics should be considered in the
phase space $$\Omega_+= R_+^n \times R^n,$$ where $R_+$ is the set of
nonnegative real numbers. We shall not study this problem in
details, because our aim is the study of the corresponding quantum
dynamics. But in the quantum case this problem is solved easily. One
should just consider the corresponding Hamiltonian in the space of
square integrable functions $L_2(\Omega_+).$ Another possibility in
the classical case is to consider centered dynamics of prices:
$$z_j(t)= q_j(t) -q(0).$$ The centered price $z_j(t)$ evolves in the
configuration space $R^n.$

\subsection{Critique of classical econophysical model}

The model of Hamiltonian price dynamics on the price phase space can
be useful to describe a market that  depends only on ``hard"
economic conditions: natural resources, volumes of production, human
resources and so on.  However, the classical price dynamics could
not be applied (at least directly) to modern financial markets. It
is clear that the stock market is not based only on these ``hard"
factors. There are other factors, soft ones (behavioral), that play
the important and (sometimes even determining) role in forming of
prices at the financial market.
 Market's psychology should be taken into account.
 Negligibly small amounts of information (due to the rapid exchange
of information) imply large changes of prices at the financial
market. We can consider a model in that financial (psychological)
waves are permanently present at the market. Sometimes these waves
produce uncontrollable changes of prices disturbing the whole market
(financial crashes). Of course, financial waves also depend on
``hard economic factors.'' However, these  factors do not play the
crucial role in forming of financial waves. Financial waves are
merely waves of information.

We could compare behavior of
financial market with behavior of a gigantic ship that is ruled by a
radio signal. A radio signal with negligibly small physical energy
can essentially change (due to information contained in this signal)
the motion of the gigantic ship. If we do not pay attention on (do
not know about the presence of) the radio signal, then we will be
continuously disappointed by ship's behavior. It can change the
direction of motion without any "hard" reason (weather, destination,
technical state of ship's equipment). However, if we know about the
existence of radio monitoring, then we could  find information that
is sent by radio. This would give us the powerful tool to predict
ship's trajectory. We now inform the reader that this example on
ship's monitoring was taken from the book of D. Bohm and B. Hiley
\cite{BOHM1} on so called pilot wave quantum theory (or Bohmian
quantum mechanics).

\section{Quantum-like Econophysical Model for Financial Market}

\subsection{Financial pilot waves}

If we  interpret the pilot wave as a field, then we should pay
attention that this is a rather strange field. It differs crucially
from ``ordinary physical fields,'' i.e., the electromagnetic field.
We mention some of the pathological features of the pilot wave field, see
\cite{BOHM}, \cite{BOHM1}, \cite{HOL} for the detailed analysis. In
particular, the force induced by this pilot wave field does not
depend on the amplitude of wave. Thus small waves and large waves
equally disturb the trajectory of an elementary particle. Such
features of the pilot wave give the possibility to speculate, see
\cite{BOHM1}, \cite{HILEY}, that this is just a wave of information
(active information). Hence, the pilot wave field  describes  the
propagation of information. The pilot wave is more similar to a
radio signal that guides a ship. Of course, this is just an analogy
(because a radio signal is related to an ordinary physical field,
namely, the electromagnetic field). The more precise analogy is to
compare the pilot wave with information contained in the radio
signal.

We remark that the pilot wave (Bohmian) interpretation of quantum
mechanics is not the conventional one. As we have already noted,
there are a few critical arguments against Bohmian quantum
formalism:

\bigskip

1. Bohmian theory gives the possibility to provide the mathematical
description of the trajectory $q(t)$ of an elementary particle.
However, such a trajectory does not exist according to
 the conventional quantum formalism.

\bigskip

2. Bohmian theory is not local, namely, via the pilot wave field one
particle "feels" another on large distances.

We say that these disadvantages of theory will become advantages in
our applications of Bohmian theory to financial market. We also
recall that already Bohm and Hiley \cite{BOHM1} and Hiley and
Pilkk\"anen\cite{HILEY} discussed the possibility to interpret the
pilot wave field as a kind of information field. This information
interpretation was essentially developed in works of Khrennikov
\cite{KHR} that were devoted to pilot wave cognitive models.

Our fundamental assumption is that agents at the modern financial
market are not just ``classical-like agents.'' Their actions are
ruled not only by classical-like financial potentials $V(t, q_1,
\ldots, q_n),$ but also (in the same way as in the pilot wave theory
for quantum systems) by an additional information (or psychological)
potential induced by a financial pilot wave.

Therefore we could not use the classical financial dynamics
(Hamiltonian formalism) on the financial phase space to describe the
real price trajectories. Information (psychological) perturbation of
Hamiltonian equations  for price and price change must be taken into
account. To describe such a model mathematically, it is convenient
to use such an object as a {\it financial pilot wave} that rules the
financial market.

In some sense  $\psi(q)$ describes the psychological influence of
the price configuration $q$ to behavior of agents of the financial
market. In particular, the $\psi(q)$ contains expectations of
agents.\footnote{  The reader may be surprised that there appeared
complex numbers $C.$ However, the use of these numbers is just a
mathematical trick that provides the simple mathematical description
of dynamics of the financial pilot wave.}

We underline two important features of the financial pilot wave
model:

\bigskip

1. All shares are coupled on the information level. The general
formalism \cite{BOHM}, \cite{BOHM1}, \cite{HOL} of the pilot wave
theory says that if the function $\psi(q_1, \ldots, q_n)$ is not
factorized, i.e.,
$$\psi(q_1, \ldots, q_n)\not = \psi_1(q_1) \ldots \psi_n(q_n),$$
then any changing the price $q_i$  will automatically change
behavior of all agents of the financial market (even those who have
no direct coupling with $i$-shares). This will imply changing of
prices of $j$-shares for $i\not= j.$ At the same time the "hard"
economic potential $V(q_1, \ldots, q_n)$ need not contain any
interaction term.   

For example, let us consider at the moment the potential 
$$V(q_1, \ldots, q_n)=q_1^2+\ldots
q_n^2.$$  The Hamiltonian equations for this potential -- in the absence of the financial
pilot wave -- have the form: $$\dot q_j=p_j, \dot p_j=-2 q_j,
j=1,2,\ldots, n.$$ Thus the classical price trajectory $ q_j(t),$
does not depend on dynamics of prices of shares for other traders
$i\not = j$ (for example, the price of shares of ERIKSSON does not
depend on the price of shares of NOKIA  and vice
versa).\footnote{Such a dynamics would be natural if these
corporations operate on independent markets, e.g., ERIKSSON in
Sweden and NOKIA in Finland. Prices of their shares would depend
only on local market conditions, e.g., on capacities of markets or
consuming activity.}

\bigskip

 However, if, for example, the wave function has the form:
$$\psi(q_1, \ldots, q_n)=c e^{i(q_1 q_2 + \ldots + q_{n-1}q_n)} e^{-(q_1^2+\ldots + q_n^2)},$$
where $c \in C$ is some normalization constant, then financial
behavior of agents at the financial market is nonlocal (see further
considerations).

\bigskip

2. Reactions of the market do not depend on the amplitude of the
financial pilot wave:  waves $\psi, 2\psi, 100000 \psi$ will produce
the same reaction. Such a behavior at the  market is quite natural
(if the financial pilot wave is interpreted as an information wave,
the wave of financial information). The amplitude of an information
signal does not play so large role in the information exchange. The
most important is the context of such a signal. The context is given
by the shape of the signal, the form of the financial pilot wave
function.

\subsection{The dynamics of prices guided by the financial pilot wave}

In fact, we need not develop a new mathematical formalism. We will
just apply the standard pilot wave formalism
 to the financial market.
The fundamental postulate of the pilot wave theory is that the pilot
wave (field) $$\psi (q_1, \ldots, q_n)$$ induces a new (quantum)
potential  $$U(q_1, \ldots, q_n)$$ which perturbs the classical
equations of motion. A modified Newton equation has the form:
\begin{equation} \label{FM}
 \dot p = f + g,
\end{equation}
where
$$
f=-\frac{\partial V}{\partial q}$$ and $$g=-\frac{\partial
U}{\partial q}.$$ We call the additional financial force $g$ a {\it
financial mental force.} This force $g(q_1, \ldots, q_n)$ determines
a kind of collective consciousness of the financial market. Of
course, the $g$ depends on economic and other `hard' conditions
given by the financial potential $V(q_1, \ldots, q_n)$. However,
this is not a direct dependence. In principle, a nonzero financial
mental force can be induced by the financial pilot wave $\psi$ in
the case of zero financial potential, $V\equiv 0.$ So $V\equiv 0$
does not imply that $U\equiv 0.$ {\it Market psychology is not
totally determined by economic factors.} Financial (psychological)
waves of information need not be generated by some changes in a real
economic situation. They are mixtures of mental and economic waves.
Even in the absence of economic waves, mental financial waves can
have a
 large influence to the  market.

By using the standard pilot wave formalism we obtain the following
rule for computing the financial mental force. We represent the
financial pilot wave $\psi(q)$ in the form:
\[\psi(q)=R(q) e^{iS(q)}\]
where $R(q)=|\psi(q)|$ is the amplitude of $\psi(q),$ (the absolute
value of the complex number $c=\psi(q)$) and $S(q)$ is the phase of
$\psi(q)$ (the argument of the  complex number $c=\psi(q)$). Then
the financial mental potential is computed as
\[U(q_1, ..., q_n) = - \frac{1}{R} \sum_{i=1}^n \frac{\partial^2R}{\partial q_i^2} \]
and the financial mental force as
\[g_j(q_1, \ldots, q_n) = \frac{-\partial U}{\partial q_j}(q_1, \ldots, q_n).\]
These formulas imply that strong financial effects are produced by
financial waves having essential variations of amplitudes.
\medskip

{\bf Example  1.} (Financial waves with small variation have no
effect). Let us start with the simplest example: $$R\equiv const.$$ Then the financial (behavioral) force
$g \equiv 0.$ As $R \equiv const,$ it is impossible to change
expectations of the whole financial market by varying the price
$q_j$ of one fixed type of shares, $j.$ The constant information
field does not induce psychological financial effects at all. As we
have already remarked the absolute value of this constant does not
play any role. Waves of constant amplitude $R=1,$ as well as
$R=10^{100},$ produce no effect.

Let now consider the case: $$R(q)=c q, c>0.$$ This is a linear function; variation is not so
large. As the result $g\equiv 0$ here also. No financial behavioral
effects.

\medskip

{\bf Example 2.} (Speculation) Let $$R(q)=c (q^2+d), \; \; c, d>0.$$
Here
$$U(q)=-\frac{2}{q^2+d}$$ (it does not depend on the amplitude $c$ !)
and $$g(q)=\frac{-4q}{(q^2+d)^2}.$$ The quadratic function varies
essentially more strongly than the linear function, and, as a
result, such a financial pilot wave induces a nontrivial financial
force.

We analyze financial drives induced by such a force. We consider the
situation: (the starting price) $q>0$  and $g<0.$ The financial
 force $g$ stimulates the  market (which works as a
huge cognitive system) to decrease the price. For small prices,
$$g(q) \approx - 4q/d^2.$$ If the financial market increases the price
$q$ for shares of this type, then the negative reaction of the
financial force becomes stronger and stronger. The  market is
pressed (by the financial force) to stop increasing of the price
$q.$ However, for large prices, $$g(q)\approx -4/q^3.$$ If the
market can approach this range of prices (despite the negative
pressure of the financial force for relatively small $q)$ then the
market will feel decreasing of the negative pressure (we recall that
we consider the financial market as a huge cognitive system). This
model explains well the speculative behavior of the financial market.

\medskip

{\bf Example 3.} 
Let now $$R(q)=c (q^4+ b),\; \;  c, b>0.$$ Thus $$g(q)=\frac{b q -
q^5}{(q^4+b)^2}.$$ Here the behavior of the market is more complicated.
Set $$d=^4\sqrt b.$$ If the price $q$ is changing from $q=0$ to $q=d$
then the market is motivated (by the financial force $g(q))$ to
increase the price. The price $q=d$ is critical for his financial
activity. By psychological reasons (of course, indirectly based on
the whole information available at the  market) the market
"understands" that it would be dangerous to continue to increase the
price. After approaching the price $q=d,$ the market has the
psychological stimuli to decrease the price.

\bigskip

Financial pilot waves $\psi(q)$ with $R(q)$ that are polynomials of
higher order can induce very complex behavior. The interval $[0,
\infty)$ is split into a collection of subintervals
$0<d_1<d_2<\ldots<d_n<\infty$ such that at each price level $q=d_j$
the trader changes his attitude to increase or to decrease the
price.

In fact, we have considered just a one-dimensional model. In the
real case we have to consider multidimensional models of huge
dimension. A financial pilot wave $\psi(q_1, \ldots, q_n)$ on such a
price space $Q$ induces splitting of $Q$ into a large number of
domains $$Q=O_1 \bigcup \ldots \bigcup O_N.$$

The only problem which we have still to solve is the description of
the time-dynamics of the financial pilot wave, $\psi(t, q).$ We
follow the standard pilot wave theory. Here $\psi(t, q)$ is found as
the solution of Schr\"odinger's equation. The Schr\"odinger equation
for the energy $$H(q,
p)=\frac{1}{2}\sum_{j=1}^{n}\frac{p_j^2}{m_j}+V(q_1, \ldots, q_n)$$
has the form:
$$
i h \frac{\partial \psi}{\partial t}(t, q_1, \ldots, q_n)=
$$
\begin{equation}
\label{SH} - \sum_{j=1}^{n} \frac{h^2}{2 m_j}\frac{\partial^2\psi(t,
q_1, \ldots, q_n)}{\partial q_j^2} + V(q_1, \ldots, q_n)\psi(t,
q_1,\ldots, q_n),
\end{equation}
with the initial condition $$\psi(0, q_1, \ldots, q_n)=\psi(q_1,
\ldots, q_n).$$  Thus if we know $\psi(0, q)$ then by using
Schr\"odinger's equation we can find the pilot wave at any instant
of time $t, \psi(t, q).$ Then we compute the corresponding mental
potential $U(t, q)$ and mental force $g(t, q)$ and solve Newton's
equation.

We shall use the same equation to find the evolution of the
financial pilot wave. We have only to make one remark, namely, on
the role of the constant $h$ in Schr\"odinger's equation, cf. E.
Haven \cite{HA2}, \cite{HA3}, \cite{HA}. In quantum mechanics (which
deals with microscopic objects) $h$ is the Planck constant. This
constant is assumed to play the fundamental role in all quantum
considerations. However, originally $h$ appeared as just a scaling
numerical parameter for processes of energy exchange. Therefore in
our financial model we can consider $h$ as a price scaling
parameter, namely, the unit in which we would like to measure price
change. We do not present any special value for $h.$ There are
numerous investigations into price scaling. It may be that there can
be recommended some special value for $h$ related to the modern
financial market, a {\it fundamental financial constant.} However,
it seems that $$h=h(t)$$ evolves depending on economic development.

We suppose that the financial pilot wave evolves via the financial
Schr\"odinger equation (an analogue of Schr\"odinger's equation) on
the price space. In the general case this equation has the form:
\[i h\frac{\partial \psi}{\partial t}(t, q)=\widehat H \psi(t,q),\psi(0, q)=\psi(q),
\]
where $\widehat H$ is self-adjoint operator corresponding to the
financial energy given by a function $H(q, p)$ on the financial
phase space. Here we proceed in the same way as in ordinary quantum
theory for elementary particles.

\subsection{On the choice of a measure of classical fluctuations}

As the mathematical basis of the model we use the space $L_2(Q)$ of
square integrable functions $\psi: Q\rightarrow {\bf C},$ where $Q$
is the configuration price space, $Q=R^n,$ or some domain $Q \subset
R^n$ (for example, $Q=R_+^n):$
$$||\psi||^2=\int_Q |\psi(x)|^2 dx<\infty.$$

Here $dx$ is the Lebesque measure, a uniform probability
distribution, on the configuration price space. Of course, the
uniform distribution $dx$ is not the unique choice of the
normalization measure on the configuration price space. By choosing
$dx$ we assume that in the absence of the pilot wave influence, all
prices ``have equal rights.'' In general, this is not true. If there
is no financial (psychological) waves the financial market still
strongly depends on ``hard'' economic conditions. In general, the
choice of the normalization measure $M$ must be justified by a real
relation between prices. So in general the financial pilot wave
$\psi$ belongs to the space $L_2(Q, d M)$ of square integrable
functions with respect to some measure $M$ on the configuration
price space:
$$||\psi||^2=\int_Q|\psi(x)|^2 d
M(x)<\infty.$$

 In particular, $M$ can be a Gaussian measure:
$$d M(x)=\frac{1}{(2\pi det B)^{n/2}}e^\frac{-(B^{-1}(x
-\alpha),x-\alpha)}{2}dx,$$ where $B=(b_{ij})^n_{i, j=1}$ is the
covariance matrix and $\alpha=(\alpha_1, \ldots, \alpha_n)$ is the
average vector. The measure $M$ describes classical random
fluctuations in the financial market that are not related to
`quantum' (behavioral) effects. The latter effects are described in
our model by the financial pilot wave. If the influence of this wave
is very small we can use classical probabilistic models; in
particular, based on the Gaussian distribution.

\section{Comparing with conventional models for the financial market}

Our model of the stocks market differs crucially from the main
conventional models. Therefore we should perform an extended
comparative analysis of our model and known models. This is not a
simple tass and it takes a lot of efforts.

\subsection{The stochastic model} Since the pioneer paper of L.
Bachelier \cite{BA}, there was actively developed various models of
the financial market based on stochastic processes. We recall that
Bachelier determined the probability of price changes $P( v(t)\leq
v)$ by writing down what is now called the Chapman-Kolmogorov
equation. If we introduce the density of this probability
distribution: $p(t,x),$ so $P( x_t\leq x)=\int_{-\infty}^x p(t,x) d
x,$ then it satisfies to the Cauchy problem of the partial
differential equation of the second order. This equation is known in
physics as Chapman's equation and in probability theory as the
direct Kolmogorov equation. In the simplest case  the underlying
diffusion process is the Wiener process (Brownian motion), this
equation has the form (the heat conduction equation):
\begin{equation}
\label{KC} \frac{\partial p(t,x)}{\partial t} = \frac{1}{2}
\frac{\partial^2 p(t,x)}{\partial x^2}.
\end{equation}
We recall again that in the  Bachelier paper \cite{BA}, $x=v$ was
the price change variable.

For a general diffusion process we have the direct Kolmogorov
equation:
\begin{equation}
\label{KC1} \frac{\partial p(t,x)}{\partial t} = \frac{1}{2}
\frac{\partial^2 }{\partial x^2}(\sigma^2(t, x) p(t,x)) -
\frac{\partial }{\partial x} (\mu(t, x) p(t,x)).
\end{equation}
This equation is based on the diffusion process
\begin{equation}
\label{KC2} dx_t = \mu(t, x_t)\, dt + \sigma(t,x_t)\, dw_t,
\end{equation}
where $w(t)$ is the Wiener process. This equation should be
interpreted as a slightly colloquial way of expressing the
corresponding integral equation
\begin{equation}
\label{KC3} x_t = x_{t_0} + \int_{t_0}^t \mu(s, x_s) ds +
\int_{t_0}^t \sigma(s,x_s) dw_s .
\end{equation}

We pay attention that Bachelier original proposal of Gaussian
distributed price changes was soon replaced by a model of in which
prices of stocks are {\it log-normal distributed}, i.e., stocks
prices $q(t)$ are performing a {\it geometric Brownian motion.} In a
geometric Brownian motion, the difference of the logarithms of
prices are Gaussian distributed.

We recall that a stochastic process $S_t$ is said to follow a
geometric Brownian motion if it satisfies the following stochastic
differential equation: \begin{equation} \label{KC4}
dS_t=u\,S_t\,dt+v\,S\,dw_t
\end{equation} where
$w_t$ is a Wiener process (=Brownian motion) and $u$ (``the
percentage drift'') and $v$ (``the percentage volatility'') are
constants. The equation has an analytic solution:
\begin{equation}
\label{KC4Q} S_t = S_0 \exp\left((u-v^2/2)t+v w_t\right)
\end{equation}
The $S_t = S_t(\omega)$ depends on a random parameter $\omega;$ this
parameter will be typically omitted. The crucial property of the
stochastic process $S_t$  is that the random variable
$$\log(S_t/S_0)= \log(S_t) - \log(S_0)$$
is normally distributed.

In the opposition to such stochastic models our Bohmian model of the
stocks market is not based on the theory stochastic differential
equations. In our model the randomness of the  stocks market
cannot be represented in the form of some transformation of the
Wiener process.

We recall that the stochastic process model was intensively
criticized by many reasons, see, e.g., \cite{MAN}.

First of all there is a number of difficult problems that could be
interpreted as technical problems. The most important among them is
the problem of the choice of an adequate stochastic process $\xi(t)$
describing price or price change. Nowadays it is widely accepted
that the GBM-model provides only a first approximation of what is
observed in real data. One should try to find new classes of
stochastic processes. In particular, they would provide the
explanation to the empirical evidence that the tails of measured
distributions are fatter than expected for a geometric Brownian
motion. To solve this problem, Mandelbrot proposed to consider the
price changes that follow a {\it Levy distribution} \cite{Man}.
However, the Levy distribution has a rather pathological property:
its variance is infinite. Therefore, as was emphasized in the book
of R. N. Mantegna and H. E. Stanley \cite{MAN}, the problem of
finding a stochastic process providing the adequate description of
the stocks market is still unsolved.

However, our critique of the conventional stochastic processes
approach to the stocks market has no direct relation to this
discussion on the choice of an underlying stochastic process. We are
more close to scientific groups which criticize  this conventional
model by questioning the possibility of describing of price dynamics
by stochastic processes at all.

\subsection{The deterministic dynamical model}

In particular, there was done a lot in applying of deterministic
nonlinear dynamical systems to simulate financial time series, see
\cite{MAN} for details. This
approach is typically criticized through the following general
argument: ``the time evolution of an asset price depends on all
information affecting the investigated asset and it seems unlikely
to us that all this information can be essentially described by a
small number of nonlinear equations,'' \cite{MAN}. We support such a
viewpoint.

We shall use only critical arguments against the hypothesis of the
stochastic stocks market which were provided by adherents of the
hypothesis of deterministic (but essentially nonlinear) stocks
market.

Only at the first sight is the Bohmian financial model is a kind of
deterministic model. Of course, dynamics of prices (as well as price
changes) are deterministic. It is described by the Newton second law,
see the ordinary differential  equation (\ref{FM}). It seems that
randomness can be incorporated into such a model only through the
initial conditions:
\begin{equation}
\label{FM2}
 \dot p(t,\omega) = f(t, q(t,\omega)) + g(t, q(t,\omega)),
 q(0)=q_0(\omega), p(0)= p_0(\omega),
\end{equation}
where $q(0)=q_0(\omega), p(0)= p_0(\omega)$ are random variables
(initial distribution of prices and momenta) and here $\omega$ is a
chance parameter.

However, the situation is not so simple. Bohmian randomness is not
reduced to randomness of initial conditions or chaotic behavior of
the equation (\ref{FM}) for some nonlinear classical and quantum
forces. These are classical impacts to randomness. But a really new
impact is given by the essentially quantum randomness which is
encoded in the $\psi$-function (=pilot wave=wave function). As we
know, the evolution of the $\psi$-function is described by an
additional equation -- Schr\"odinger's equation -- and hence the
$\psi$-randomness could be extracted neither from the initial
conditions for (\ref{FM2}) nor from possible chaotic behavior.

In our model the $\psi$-function gives the dynamics of expectations
at the financial market. These expectations are a huge source of
randomness at the market -- mental (psychological) randomness.
However, this randomness is not classical (so it is a non-Kolmogorov
probability model).

Finally, we pay attention that in quantum mechanics the wave
function is not a measurable quantity. It seems that a similar
situation we have for the financial market. We are not able to
measure the financial $\psi$-field (which is an infinite dimensional
object, since the Hilbert space has the infinite dimension). This
field contains thoughts and expectations of millions agents and of
course it could not be ``recorded'' (in the opposition to prices or
price changes).

\subsection{The stochastic model and expectations of the agents of
the financial market}

Let us consider again the model of the stocks market based on the
geometric Brownian motion:
$$
dS_t=u\,S_t\,dt+v\,S\,dw_t.
$$
We pay attention that in this equation there is no term describing
the behavior of agents of the  market.  Coefficients $u$ and $v$ do
not have any direct relation to expectations and the market
psychology. Moreover, if we even introduce some additional
stochastic processes
$$\eta(t,\omega)= (\eta_1(t,\omega),..., \eta_N(t,\omega)).$$
describing behavior of agents and additional coefficients (in
stochastic differential equations for such processes) we would be
not able to simulate the real market. A finite dimensional vector
$\eta(t,\omega)$ cannot describe the ``mental state of the market''
which is of the infinite complexity. One can consider the Bohmian
model as the introduction of the infinite-dimensional chance
parameter $\psi.$ And this chance parameter cannot be described by
the classical probability theory.

\subsection{The efficient market hypothesis and the Bohmian approach
to financial market}

The efficient market hypothesis was formulated in sixties, see
\cite{SM} and \cite{Fam} for details:

{\it A market is said to be efficient in the determination of the
most rational price if all the available information is instantly
processed when it reaches the market and it is immediately reflected
in a new value of prices of the assets traded.}

The efficient market hypothesis is closely related to the stochastic
market hypothesis. Mathematically the efficient market hypothesis
was supported by investigations of Samuelson \cite{SM}. Using the
hypothesis of rational behavior and market efficiency he was able to
demonstrate how $q_{t+1},$  the expected value of price of a given
asset at time $t+1,$ is related to the previous values of prices
$q_0, q_1,..., q_t$ through the relation
\begin{equation}
\label{SM} E (q_{t+1}\vert q_0, q_1,..., q_t)= q_t.
\end{equation}
Stochastic processes of such a type are called martingales
\cite{Sh}.

Thus the efficient market hypothesis implies that the financial
market is described by a special class of stochastic processes -
martingales, see A. Shiryaev \cite{Sh}.

Since the Bohmian quantum model for the financial market is not
based on the the stochastic market hypothesis, the efficient market
hypothesis can be neither used as the basis of the Bohmian quantum
model. The relation between the efficient market model and the
Bohmian quantum market model is very delicate. There is no direct
contradiction between these models. Since classical randomness is
also incorporated into the Bohmian quantum market model (through
randomness of initial conditions), we should agree that ``the
available information is instantly processed when it reaches the
market and it is immediately reflected in a new value of prices of
the assets traded.'' However, besides the available information
there is information encoded through the $\psi$-function describing
the market psychology. As was already mentioned, $\psi$-function is
not measurable, so the complete information encoded in this function
is not available. Nevertheless, some parts of this information can
be extracted from the $\psi$-function by some agents of the
financial market (e.g., by those who ``feel better the market
psychology''). Therefore classical forming of prices based on the
available information is permanently disturbed by quantum
contributions to prices of assets. Finally, we should conclude that
the real financial market (and not it idealization based on the
mentioned hypothesizes) is not efficient. In particular, it
determine not the most rational price. It may even induce completely
irrational  prices through quantum effects.

\section{On Views of G. Soros: Alchemy of Finances or Quantum Mechanics of Finances?}

G. Soros is unquestionably the most powerful and profitable investor
in the world today. He has made a billion dollars going against the
British pound.

Soros is not merely a man of finance, but a thinker and philosopher
as well. Surprisingly he was able to apply his general philosophic
ideas to financial market. In particular, the project Quantum Fund
inside {\it Soros Fund Management} gained hundreds millions dollars
and has 6 billion dollars in net assets.

The book ``Alchemy of Finance'' \cite{S} is a kind of
economic-philosophic investigation. In my thesis I would like to
analyze philosophic aspects of this  investigation.  The book
consists of five chapters. In fact, only the first chapter -
``Theory of Reflexivity''  - is devoted to pure theoretical
considerations.

J. Soros studied economics in college, but found that economic
theory was highly unsatisfactory. He says that economics seeks to be a
science, but science is supposed to be objective. And it is difficult
to be scientific when the subject matter, the participant in the
economic process, lacks objectivity. The author also was greatly
influenced  by Karl Popper's ideas on scientific method, but he did
not agree with Popper's  ``unity method.'' By this Karl Popper meant
that methods and criteria which can be applied to the study of
natural phenomena  also can be applied to the study of social
events. George Soros underlined a fundamental difference between
natural and social sciences:

The events studied by social sciences  have thinking participants
and natural phenomena do not. The participants' thinking creates
problems that have no counterpart in natural science. There is a
close analogy with QUANTUM PHYSICS, where the effects of scientific
observations give rise to Heisenberg uncertainty relations and
Bohr's complementarity principle.

But in social events the participants' thinking is responsible for
the element of uncertainty, and not an  external observer. In natural
science investigation of events goes from fact to fact. In social
events the chain of causation does not lead directly from fact to
fact, but from fact to participants' perceptions and from
perceptions to fact.

This would not create any serious difficulties if there were
some kind of correspondence or equivalence between facts and
perceptions.

Unfortunately, that is impossible, because the participants
perceptions do not relate to facts,  but to a situation  that is
contingent on their own perceptions and therefore cannot be treated
as a fact.

In order to appreciate the problem posed by thinking participants,
Soros takes a closer look at the way scientific method operates. He
takes Popper's scheme of scientific method, described in technical
terms as ``deductive-nomological'' or ``D-N'' model. The model is
built on three kinds of statements: specific initial conditions,
specific final conditions, and generalizations of universal
validity. Combining a set of generalizations with known initial
conditions yields predictions, combining them with known final conditions
provides explanations; and matching known initial with known final
conditions serves as testing for generalizations involved.
Scientific theories can only be falsified, never verified.

The asymmetry between verification and falsification and the
symmetry between prediction and explanation are two crucial features
of Popper's scheme.

The model works only if certain conditions are fulfilled. It is the
requirement of universality. That is, if a given set of conditions
recurred, it would have to be followed or predicted by the same set
of conditions as before. The initial and final conditions must
consist of observable facts governed by universal laws. It is this
requirement that is so difficult to meet when a situation has
thinking participants. Clearly, a single observation by a single
scientist is not admissible. Exactly because the correspondence
between facts and statements is so difficult to establish, science
is a collective enterprize  where the work of each scientist has to
be open to control and criticism by others. Individual scientists
often find the conventions quite  onerous and try various shortcuts
in order to obtain a desired result. The most outstanding example of
the observer trying to impose his will  on his subject matter is the
attempt to convert base metal into gold. Alchemists  struggled long
and hard until they were finally persuaded to abandon their
enterprize by their lack of success. The failure was inevitable
because the behavior of base metals is governed by laws of universal
validity which cannot be modified by any statements, incantations,
or rituals.

And now, we can at least understand why J. Soros called his book
{\it Alchemy of Finance}, see \cite{S}.

Soros considers the behavior of human beings. Do they obey
universally valid laws that can be formulated in accordance with
``D-N'' model? Undoubtedly, there are many aspects of human
behavior, from birth to death and in between, which are amenable to
the same treatment as other natural phenomena. But there is one
aspect of human behavior which seems to exhibit characteristics
which are different from those of the phenomena from the subject
matter of natural science: the decision making process. An imperfect
understanding of the situation destroys the universal validity of
scientific generalizations: given a set of conditions is not
necessary preceded or succeeded by the same set every time, because
the sequence of events is influenced by participants' thinking. The
``D-N'' model breaks down. But social scientists try to maintain the
unity of  method but with little success.

In a sense, the attempt to impose the methods of natural science on
social phenomena is comparable to efforts of alchemists who sought
to apply the methods of magic to the field of natural science. And
here J. Soros presents (in my opinion) a very interesting idea about
the ``alchemy method in social science.'' He says, that while the
failure of the alchemists was total, social scientists have managed
to make a considerable impact on their subject matter. Situations
which have thinking participants may be impervious  to the methods
of natural science, but they are susceptible to the methods of
alchemy.

The thinking of participants, exactly because it is not governed by
reality, is easily influenced by theories. In the field of natural
phenomena, scientific method is effective only when its theories are
valid, but in social, political, and economic matters, theories can
be effective without being valid. Whereas  alchemy has failed in
natural sciences, social science can succeed as alchemy.

The relationship between the scientist and his subject matter is quite 
different in natural science as opposed to social science. In natural science the
scientist's thinking is, in fact, distinct from its subject matter.
The scientists can influence the subject matter only by actions, not
by thoughts, and the scientists actions are guided by the same laws
as all other natural phenomena. Specifically, a scientist can do nothing
do when she wants to turn base metals into gold.

Social phenomena are different. The imperfect understanding of the
participant interferes with  the proper functioning of the ``D-N''
model. There is much to be gained by pretending to abide by
conventions of scientific method without actually doing so. Natural
science is held in great esteem: the theory that claims to be
scientific can influence the gullible public much better than one
which frankly admits its political or ideological bias.

Soros mentions here Marxism, psychoanalysis and laissez-faire
capitalism with its reliance on the theory of perfect competition as
typical examples.

Soros underlined that Marx and Freud were vocal in protesting their
scientific status and based many of their conclusions on authority they
derived from being ``scientific.'' And Soros says, that once this
point sinks in, the very expression ``social science'' became
suspect.

He compares the expression ``social science'' with a magic word
employed by social alchemists in their effort to impose their will
on their subject matter by incantation. And it seems to Soros there is only
one way for the ``true'' practitioners of scientific method to protect
themselves against such malpractice - to deprive social science of
the status it enjoys on account of natural science. Social science
ought to be recognized  as a false metaphor.

I cannot agree with this Soros statement. First of all there are a
lot of boundary sciences. For example, psychoanalysis can be
considered as a part of medicine. But medicine accompanied with
biology and chemistry are of course the natural sciences. And S.
Freud was famous and like many doctors succeeding him they helped many patients.

And J. Soros explains by himself his unusual  statement. He says
that it  does not mean that we must give up the pursuit of truth in
exploring social phenomena. It means only that the pursuit of truth
requires to recognize that the ``D-N'' model can not be applied to
situations with thinking participants. He asks us to abandon the
doctrine of the unity of method and to cease the slavish imitation of
natural sciences. He says that there are some new scientific methods
in all kinds of science as quantum physics has shown.

Scientific method is not necessarily  confined to the ``D-N'' model:
statistical, probabilistic generalizations may be more fruitful. Nor
should we ignore the possibility of developing novel approaches
which have no counterpart in natural science. Given the differences
in subject matter, there ought to be differences in the method of
study.

Soros shows us the main distinction between the ``D-N'' model and his
own approach. He says that the world of imperfect understanding does
not land itself to generalizations which can be used to explain and
to predict specific events. The symmetry between explanation and
prediction prevails only in the absence of thinking participants. On
the other hand, past events are just as final as in the ``D-N''
model; thus explanations turn out to be an easier task than
prediction.

In another part of his book , see \cite{S}, Soros abandons the
constraint that predictions and explanations are logically
reversible. He builds his own theoretical framework. He says that
his theory can not be tested in the same way as these theories which fit into
Popper's logical structure, but that is not to say that testing must
be abandoned. And he does tests in a real-time experiment,
chapter 3, for his model. He uses the theory of perfect competition
for the investigation of financial market, but he takes into account
thinking of participants of this market.

G. Soros proved his theory by becoming one of the most powerful and profitable
investors in the world today. In my own work I use methods of
classical and quantum mechanics for mathematical modeling of price
dynamics at financial market and  I use Soros' statement about
cognitive phenomena at the financial market.

\section{Existence Theorems for Non-smooth Financial Forces}

\subsection{The problem of smoothness of price trajectories}

In the Bohmian model for price dynamics the price trajectory $q(t)$
can be found as the solution of the equation
\begin{equation}
\label{LM} m \frac{d^2 q(t)}{dt^2}=f(t, q (t)) + g (t, q(t))
\end{equation}
with the initial condition
$$
q(t_0)= q_0, q^\prime (t_0)= q_0^\prime.
$$
Here we consider a "classical" (time dependent) force $$f(t, q)=-
\frac{\partial V(t, q)}{\partial q}$$ and "quantum-like" force $$g(t, q)=-
\frac{\partial U(t, q)}{\partial q},$$ where $U(t, q)$ is the quantum
potential, induced by the Schr\"odinger dynamics. In Bohmian
mechanics for {\it physical systems} the equation (\ref{LM}) is
considered as an ordinary differential equation and $q(t)$ as the
unique solution (corresponding to the initial conditions
$q(t_0)=q_0, q^\prime (t_0)= q_0^\prime)$ of the class $C^2: q(t)$
is assumed to be twice differentiable with continuous
$q^{\prime\prime} (t).$

One of possible objections to  apply the Bohmian quantum model to
describe dynamics of prices (of e.g. shares) at the financial market
is smoothness of trajectories. In financial mathematics it is
commonly assumed that the price-trajectory is not differentiable,
see, e.g., \cite{MAN}, \cite{Sh}.

\subsection{Mathematical model and reality}

Of course, one could simply reply that there are no smooth
trajectories in nature. Smooth trajectories belong neither to
physical nor financial reality. They appear in mathematical models
which can be used to describe reality. It is clear that the
possibility to apply a mathematical model with smooth trajectories
depends on a chosen time scale. Trajectories that can be considered
as smooth (or continuous) at one time scale might be nonsmooth (or
discontinuous) at a finer time scale.

We illustrate this general philosophic thesis by the history of
development of financial models. We recall that at the first stage
of development of financial mathematics, in the Bachelier model and
the Black and Scholes model, {\it there were considered processes
with continuous trajectories:} the Wiener process and more general
diffusion processes. However, recently it was claimed that such
stochastic models (with continuous processes) are not completely
adequate to real financial data, see, e.g., \cite{MAN}, \cite{Sh}
for the detailed analysis. It was observed that at finer time scales
some Levy-processes with jump-trajectories are more adequate to data
from the financial market.

Therefore one could say that the Bohmian model provides a rough
description of price dynamics and describes not the real price
trajectories by their smoothed versions. However, it would be
interesting to keep the interpretation of Bohmian trajectories as
the real price trajectories. In such an approach one should obtain
nonsmooth Bohmian trajectories. The following section is devoted to
theorems providing existing of nonsmooth solutions.

\subsection{Picard's theorem and its generalization}

We recall the standard  uniqueness and existence theorem for
ordinary differential equations, Picard's theorem, that gives the
guarantee of smoothness of trajectories, see, e.g., \cite{KOLF}.

\medskip

{\bf Theorem 1.} {\it Let $F: [0, T] \times {\bf R} \to {\bf R}$ be
a continuous function and let $F$ satisfy the Lipschitz condition
with respect to the variable $x$:
\begin{equation}
\label{LM1} |F(t, x)-F(t, y)|\leq c |x-y|, c > 0.
\end{equation}
Then, for any point ($t_0, x_0) \in [0, T) \times {\bf R}$ there
exists the unique $C^1$-solution of the Cauchy problem:
\begin{equation}
\label{LM2} \frac{dx}{dt}= F(t, x(t)), \; \; x(t_0)= x_0,
\end{equation}
on the segment $\Delta=[t_0, a],$ where $a>0$ depends $t_0, x_0,$
and $F.$}

\medskip

We recall the standard proof of this theorem, because the scheme of
this proof will be used later. Let us consider the space of
continuous functions $x: [t_0, a] \to {\bf R},$ where $a>0$ is a
number which will be determined. Denote this space by the symbol
$C[t_0, a].$ The Cauchy problem (\ref{LM2}) for the ordinary
differential equation can be written as the integral equation:
\begin{equation}
\label{LM3} x(t)=x (t_0) + \int_{t_0}^t F(s, x(s)) ds
\end{equation}
The crucial point for our further considerations is that continuity
of the function $F$ with respect to the pair of variables $(t, x)$
implies continuity of $y(s)=F(s, x(s))$ for any continuous $x(s).$
But the integral $z(t)=\int_0^t y(s) ds$ is differentiable for any
continuous $y(s)$ and $z^\prime (t)=y(t)$ is also continuous. The
basic point of the standard proof is that, for a sufficiently small
$a > 0,$ the operator
\begin{equation}
\label{LM4} G(x)(t)= x_0 + \int_{t_0}^t F(s, x (s)) ds
\end{equation}
maps the functional space $C[t_0, a]$ into $C[t_0, a]$ and it is a
contraction in this space:
\begin{equation}
\label{LM5} \rho_\infty (G(x_1),G(x_2)) \leq \alpha \rho_\infty
(x_{10}, x_{20}), \; \; \alpha < 1,
\end{equation}
for any two trajectories $x_1(t), x_2(t) \in  C[t_0, a]$ such that
$x_1(t_0)= x_{10}$ and $x_2(t_0)= x_{20}.$ Here, to obtain $\alpha <
1,$  the interval $[t_0, a]$ should be chosen sufficiently small,
see further considerations. Here $\rho_\infty (u_1, u_2)= ||u_1 -
u_2||_\infty$ and $$||u||_\infty= \sup_{t_0 \leq t \leq a}|u(s)|.$$
The contraction condition, $\alpha < 1,$ implies that the iterations
$$x_1 (t)= x_0 + \int_{t_0}^t F (S, x_0) ds,$$
$$x_2 (t)= x_0 + \int_{t_0}^t F (S, x_1 (S)) ds,...,$$
$$x_n (t)= x_0 + \int_{t_0}^t F (S, x_{n-1} (S)) ds, ...$$
converge to a solution $x(t)$ of the integral equation (\ref{LM3}).
Finally, we remark that the contraction condition (\ref{LM5})
implies that the solution is unique in the space $C[t_0, a].$

\medskip

Roughly speaking in Theorem 1 the Lipschits condition is
"responsible" for  uniqueness of solution and continuity of $F(t,
x)$ for existence. We also recall the well known Peano theorem,
\cite{KOLF}:

\medskip

{\bf Theorem 2.} {\it Let $F: [0, T] \times {\bf R}$ be a continuous
function. Then, for any point ($t_0, x_0) \in [0, T] \times {\bf R}$
there exists locally a $C^1$-solution of the Cauchy problem
(\ref{LM2}).}

\medskip

We remark that Peano's theorem does not imply uniqueness of
solution.

\medskip

 It is clear that discontinuous financial forces can
induce price trajectories $q(t)$ which are not smooth: more over,
price trajectories can even be discontinuous! From this point of
view the main problem is not smoothness of price trajectories $q(t)$
(and in particular the zero covariation for such trajectories), but
the absence of an existence and uniqueness theorem for discontinuous
financial forces. We shall formulate and prove such a theorem. Of
course, outside the class of smooth solutions one could not study
the original Cauchy problem for an ordinary differential equation
(\ref{LM2}). Instead of this one should consider the integral
equation (\ref{LM3}).

We shall generalize  Theorem 1 to discontinuous $F.$ Let us consider
the space $BM[t_0, a]$ consisting of bounded measurable functions
$x:[t_0, a] \to {\bf R}.$ Thus:

\medskip

a) $\sup_{t_0 \leq t \leq a}|x(t)|\equiv||x||_\infty  < \infty ;$

\medskip

b) for any Borel subset $A \subset {\bf R},$ its preimage
$x^{-1}(A)=\{s \in [t_0, a]: x(s) \in A\}$ is again a Borel subset
in $[t_0, a].$

\medskip

{\bf Lemma 1.} {\it The space of trajectories $BM [t_0, a]$ is a
Banach space}

{\bf Proof.} Let $\{x_n (t)\}$ be a sequence of trajectories that is
a Cauchy sequence in the space $BM [t_0, a]:$ $$||x_n - x_m||_\infty \to 0, n, m
\to \infty.$$ Thus
\begin{equation}
\label{z1} \lim_{n, m \to \infty} \sup_{t_0 \leq t \leq a} |x_n(t) -
x_m (t)|\to 0.
\end{equation}
Thus, for any $t \in [t_0, a], |x_n (t)-x_m (t)|\to 0, n, m \to
\infty.$ Hence, for any $t,$ the sequence of real numbers $\{x_n
(t)\}_{n=1}^\infty$ is a Cauchy sequence in ${\bf R}.$

But the space ${\bf R}$ is complete. Thus, for any $t \in [t_0, a],$
there exists $\lim_{n \to \infty} x_n (t),$ which we denote by
$x(t).$ In this way we constructed a new function $x(t), t \in [t_0,
a].$ We now write the condition (\ref{z1}) through the
$\epsilon$-language: $\forall \epsilon > 0 \exists N:$ $\forall n, m
\geq N:$
\begin{equation}
\label{z2} |x_n(t)-x_m (t)| \leq \epsilon \; \mbox{for any} \; t \in
[t_0, a].
\end{equation}
We now fix $n \geq N$ and take the limit $m \to \infty$ in the
inequality (\ref{z2}). We obtain:
\begin{equation}
\label{z3} |x_n (t)-x(t)|\leq \epsilon, \; {\mbox{for any}} \;  t
\in [t_0, a].
\end{equation}
Thus
\begin{equation}
\label{z4} \sup_{t_0 \leq t \leq a}|x_n (t)-x(t)|\leq \epsilon
\end{equation}
This is nothing else than the condition: $$\forall n \geq N:
||x_n-x||_\infty \leq \epsilon.$$ Therefore $x_n \to x$ in the space
$BM[t_0, a].$ We remark that the trajectory $x(t)$ is bounded,
because: $$||x||_\infty \leq ||x_{n_0}-x||_\infty +
||x_{n_0}||_\infty \in \epsilon + ||x_{n_0}||_\infty$$ for any fixed
$n_0 \geq N,$ and since $||x_{n_0}||_\infty < \infty,$ we finally
get $||x||_\infty < \infty.$ We also remark that $x(t)$ is a
measurable function as the unifrom limit of measurable functions,
see [3]. Thus the space $BM[t_0, a]$ is a complete normed space - a
Banach space.

\medskip

{\bf Theorem 3.} {\it Let $F: [0, T] \times  {\bf R} \to {\bf R}$ be
a measurable bounded function and let $F$ satisfy the Lipschitz
condition with respect to the $x$-variable, see (\ref{LM1}). Then,
for any point $(t_0, x_0 \in [0, T) \times {\bf R},$ there exists
the unique solution of the integral equation (\ref{LM3}) of the
class $BM[t_0, a],$ where $a>0$ depends on $x_0, t_0,$ and $F.$}

{\bf Proof.} We shall determine $a > 0$ later. Let $u(s)$ be any
function of the class $BM [t_0, a].$ Then the function $y(s)=F(s, u
(s))$ is measurable (since $F$ and $u$ are measurable) and it is
bounded (since $F$ is bounded). Thus $y \in BM [t_0, a].$ Any
bounded and measurable function is integrable with respect to the
Lebesque measure $dt$ on $[t_0, a],$ see [3]. Therefore
$$
z(t)=\int_{t_0}^t y (s)ds \equiv \int_{t_0}^t F (s, u(s)) ds
$$ is well defined for any $t$. This function is again measurable
with respect to $t,$ see [5], and bounded, because:
$$
\Big|\int_{t_0}^t y (a) ds\Big| \leq \sup_{t_0 \leq s \leq
t}|y(s)|(t-t_0) \leq ||y||_\infty (a-t_0) < \infty.
$$
Thus the operator $G$ which was defined by (\ref{LM4}) maps $BM[t_0,
a]$ into $BM[t_0, a].$ We now show that, for a sufficiently small $a
> 0, G$ is a contraction in $BM[t_0, a].$ By using the Lipschitz
condition we get:
$$
\sup_{t_0 \leq t \leq a}|G(x_1)(t)-G(x_2)(t)|=|\int_{t_0}^t (F(s,
x_1(s)))-F(s, x_2(s)) ds|
$$
$$
 \leq \sup_{t_0 \leq t \leq a}
\int_{t_0}^t|F(s, x_1(s))- F(s, x_2(s))|ds \leq \sup_{t_0 \leq t
\leq a} c \int_{t_0}^t|x_1(s)- x_2(s)|ds
$$
$$
\leq \sup_{t_0 \leq t \leq a} c (t-t_0) \sup_{t_0\leq s \leq
t}|x_1(s)-x_2(s)| \leq c(a-t_0)||x_1-x_2||_\infty.
$$
We set $\alpha=c(a-t_0).$ If $\alpha < 1, $ i.e., $ c(a-t_0)< 1, $
or $(a-t_0)<\frac{1}{2},$ or $0 < a < \frac{1}{c} + t_0,$ then $G$
is a contraction. Therefore by the well known fixed point for
contracting maps in complete metric spaces (in particular in Banach
spaces), see [3], the map $G$ has the unique fixed point, $$x(t) \in
BM [t_0, a], G (x)=x,$$ or $$x(t)=x_0 + \int_{t_0}^t F(s, x (s)) ds.$$

\medskip

{\bf Proposition 1.} (Continuity of the solution of the integral
equation). {\it Let conditions of Theorem 3 holds. Then solutions
are continuous functions $x:[t_0, a] \to {\bf R}$}

{\bf Proof.} We use the same notations as in proof of Theorem 3. Let
$u \in BM [t_0, a], y(s)=F(s, u(s)).$ As we have shown, this is a
bounded measurable function. We prove that $u(t)=\int_{t_0}^t y(s)
ds$ is a continuous function. Let $\tau \in [t_0, t]$ and let
$\Delta$ be a small real number. Then $$|\xi(\tau +
\Delta)-\xi(\tau)|=\Big|\int_\tau^{\tau + \Delta} y (s)
ds\Big|\leq|\Delta|||y||_\infty \to 0, \Delta \to 0.$$ Here we have
used simple properties of the  Lebesque integral: $|\int_a^e y(s)
ds|\leq \int_a^b|y(s)|ds$ and, if $|y(s)|\leq const,$ then
$\int_a^b|y(s)|ds \leq const (b-a)$ (in our case
$const=||y||_\infty=\sup_{t_0 \leq t \leq a}|y(\in)|)$.

\medskip

Thus Theorem 3  gives a sufficient condition of the existence of the
unique continuous trajectory-solution $x(t)$ for the integral
equation (\ref{LM3}). But, of course, in general $x(t)$ is not
continuously differentiable!

{\bf Theorem 4.} {\it{Let $f$ satisfy the Lipschitz condition
(\ref{LM1}). Then for any point $(t_0, x_0 \in [0,T) \times R)$
there exists the unique solution of the integral equation
(\ref{LM3}) of the class $L_2 [t_0, a], $ where $a > 0$ depends on
$x_0, t_0,$ and $F.$}}

{\bf Proof.} Let $u \in L_2 [t_0, a]$ (as always we shall  determine
$a > 0$ later). Then $y(s)=F(s, u (s))$ also belongs to the class
$L_2 [t_0, a]:$

$$\int_{t_0}^a y^2 (s) ds= \int_{t_0}^a F^2 (s, u(s)) ds$$

$$\leq \int_{t_0}^a (m_1|u(s)|+ m_2)^2 ds=$$

$$m_1^2 \int_{t_0}^a u^2(s) ds + m_2^2 (a-t_0) + 2 m_1 m_2 \int_{t_0}^a |u(s)|ds.$$ Here we estimated $F(t, u)$ through the inequality (\ref{z2}).

Now we recall the well known Cauchy-Bunyakowsky inequality in the
$L_2$ space. For any pair of trajectories $u_1, u_2 \in L_2,$ we
have
$$\int_{t_0}^a|u_1 (s) u_2 (s)|ds \leq \sqrt{\int_{t_0}^a u_1^2
(s) ds} \; \; \sqrt{\int_{t_0}^a u_2^2 (s) ds}.$$ We would like to
estimate the integral
$$\int_{t_0}^a|u(s)|ds$$
by using the Cauchy-Bunaykovsky inequality. We choose $u_2(s)= u(s)$
and $u_1(s)\equiv 1.$ We have

$$\int_{t_0}^a|u(s)|ds \leq \sqrt{\int_{t_0}^a ds}  \; \; \sqrt{\int_{t_0}^a u^2(s) ds}= \sqrt{a-t_0} \; \; ||u||_2.$$
Finally, we get

$$\int_{t_0}^a y^2(s) ds \leq m_1^2||u||_2^2 + m_2^2 (a-t_0) + 2 m_1 m_2 \sqrt{a-t_0})||u||_2 <
\infty.
$$
Thus the function $y \in L_2 [t_0, a].$ Therefore the integral
operator given by
$$
G (u) (t)=x_0 + \int_{t_0}^t F(s, u(s))ds
$$
maps the space of trajectories $L_2 [t_0, a]$ into $L_2 [t_0, a].$
We recall that $L_2$-spaces are Banach spaces.  Hence, these are
complete metric spaces. Here we can apply the fixed point theorem
for compression-maps. Finally, we shall prove that the integral
operator $G: L_2 [t_0, a] \to L_2 [t_0, a]$ is compression for a
sufficiently small $a > 0.$

As always, we use the Lipschitz condition with respect to $x$. For
any pair of trajectories $x_1(s), x_2 (s) \in L_2 [t_0, a]:$
$$
||G(x_1)-G(x_2)||_2^2= \int_{t_0}^a \Big(\int_{t_0}^t (F(S, x_1 (S))
- F(s, x_2(s))) ds\Big)^2 dt
$$
$$
\leq c^2 \int_{t_0}^a \Big(\int_{t_0}^t|x_1 (s)-x_2 (s)|ds\Big)^2
dt.
$$
We now introduce the characteristic function of the interval $[t_0,
t]:$
 \[ \phi_t(s)= \left\{ \begin{array}{ll}
 1, s \in [t_0, t]\\
0, s \not \in [t_0, t] \end{array}
 \right .
 \]
The last integral can be written as
$$
\int_{t_0}^a (\int_{t_0}^t|x_1 (s)-x_2 (s)|ds)^2 dt=  \int_{t_0}^a
\Big(\int_{t_0}^a \phi_t (s)|x_1(s)-x_2 (s)|ds\Big)^2 dt.
$$
We now apply the Cauchy-Bunaykovsky inequality for the integral with
respect to $ds.$ We choose $u_1(s)=\phi_t(s)$ and
$u_2(s)=|x_1(s)-x_2(s)|.$ We have:
$$\int_{t_0}^a \phi_t (s)|x_1(s) - x_2(s)|ds$$
$$\leq \sqrt{\int_{t_0}^a \phi_t^2 (s) ds}  \; \; \sqrt{\int_{t_0}^a|x_1(s) - x_2(s)|^2 ds}$$
$$=\sqrt{\int_{t_0}^t ds}  \; \; ||x_1 - x_2||_2= \sqrt{t-t_0}  \; \; || x_1 - x_2||_2 \leq \sqrt{a-t_0} \; \; ||x_1-x_2||_2.$$
We have, finally,
$$||G(x_1)-G(x_2)||_2^2 \leq c^2 \int_{t_0}^a (a-t_0) ||x_1-x_2||_2^2 dt \leq c^2 (a-t_0)^2 \; \; ||x_1-x_2||^2_2.$$
Thus
$$\rho_2(G(x_1), G(x_2))=||G(x_1)-G(x_2)||_2 \leq c(a-t_0)
\rho_2 (x_1, x_2).$$ We set $$\alpha=c(a-t_0).$$ Hence, if $\alpha <
1,$ then
$$
G:L_2 [t_0, a] \to L_2 [t_0, a]
$$
is a compression. It has a fixed point which is the unique solution
of our integral equation. Thus the proof is finished.

\bigskip

We remark that in the same way as in the case $BM [t_0, a]$-space,
we can show that solutions existing due to Theorem 4 are continuous
functions.

\bigskip

{\bf Proposition 2.} (Continuity) {\it Let conditions of Theorem 4
hold. Then solutions $x: [t_0, a] \to {\bf R}$ are continuous
functions.}

{\bf Proof.} As we have seen  in the proof of Theorem 4, for any
trajectory $u \in L_2[t_0, a],$ the function  $y (s)=F(s, u(s))$
also belongs to $L_2[t_0, a].$ We shall prove that
$$
\xi(s)=\int_{t_0}^t y(s)ds
$$ is continuous. Let us take $\Delta \geq
0$ (the case $\Delta < 0$ is considered in the same way). We have
$$
|\xi(\tau + \Delta) - \xi(\tau)| \leq \int_\tau^{\tau +
\Delta}|y(s)|ds.
$$
We introduce the characteristic functions
 \[ \phi_{[\tau, \tau + \Delta]}(s)= \left\{ \begin{array}{ll}
 1, s \in [\tau, \tau + \Delta] \\
 0, s \not \in [\tau, \tau + \Delta]
\end{array}
 \right.
 \]
We have: $$\int_\tau^{\tau + \Delta}|y(s)|ds= \int_{t_0}^a
\phi_{[\tau, \tau + \Delta]} (s) |y(s)|ds
$$
$$
 \leq
\sqrt{\int_{t_0}^a \phi^2_{[\tau, \tau + \Delta]}(s) ds}  \; \;
\sqrt{\int_{t_0}^a |y(s)|^2 ds} = \sqrt{\Delta} \; \; ||y||_2 \to 0,
\Delta \to 0.
$$
Here we have used the Cauchy-Bunaykovsky inequality for functions
$u_1(s)=\phi_{[\tau, \tau + \Delta]}(s)$ and $u_2(s)=|y(s)|.$ The
proof is completed.

\bigskip

Thus we again obtained continuous, but in general non-smooth $(x
\not\in C^1)$ solutions of the basic integral equation.

The theory  can be naturally generalized to $L_p$ spaces, $p \geq
1:$
$$
L_p [t_0, a]=\{x: [t_0, a] \to {\bf R}: ||x||_p^p \equiv
\int_{t_0}^a |x(t)|^p dt < \infty\}.
$$
We shall not do this, because our aim was jut to show that the
integral equation (\ref{LM3}) with discontinuous $F$ is well posed
(i.e., it has the unique solution) in some classes of (nonsmooth)
trajectories.

It is more important for us to remark that Theorems 3, 4 are valid
in the multidimensional case:
$$
x_0=(x_{01}, \ldots, x_{0n}), x(t)= (x_1 (t), \ldots, x_n (t)),
$$
and
$$F: [0, T] \times {\bf R}^n \to {\bf R}^n.$$

To show this, we should change in all previous considerations the
absolute value $|x|$ to be norm on the Euclidean space ${\bf R}^n:$
$$\Vert x \Vert =\sqrt{\sum_{j=1} x_j^2}.$$
We now use the standard trick to apply our theory to the Newton
equation (\ref{LM}) which is a second order differential equation.
We rewrite this equation as a system of equations of the first order
with respect to $$x=(x_1, \ldots, x_n, x_{n+1}, x_{2n}),$$ where
$$
x_1=q_1, \ldots, x_n=q_n,$$
$$ x_{n+1}= p_1, \ldots, x_{2n}=p_n.
$$
In fact, this is nothing else than the phase space representation.
The Newton equation (\ref{LM}) will be written as the Hamilton
equation, see section? However, the Hamiltonian structure is not
important for us in this context. In any event we obtain the
following system of the first order equations:
\begin{equation}
\label{ZY} \frac{dx}{dt}=F (t, x(t)),
\end{equation}
where
 \[F(t, x)= \left( \begin{array}{ll}
 x_{n+1}\\
\cdot\\
\cdot\\
\cdot\\
x_{2n}\\
f_1 (t, x_1, \ldots, x_n) + g_1 (t, x_1, \ldots, x_n)\\
\cdot\\
\cdot\\
\cdot\\
f_n(t, x_1, \ldots, x_n) + g_n(t, x_1, \ldots, x_n)
\end{array}
 \right ).
 \]
Here $$f_j (t, x_1, x_n)=\frac{\partial V}{\partial x_j} (t, x_1,
\ldots, x_n)
$$ and
$$g_j(t, x_1, \ldots, x_n)=\frac{\partial
U}{\partial x_j} (t, x_1, \ldots, x_n).
$$
Therefore if
$$\nabla V=\Big(\frac{\partial V}{\partial_{x_n}}, \ldots, \frac{\partial V}{\partial_{x_n}}\Big)$$ or
$$\nabla U=\Big(\frac{\partial U}{\partial_{x_n}}, \ldots, \frac{\partial U}{\partial_{x_n}}\Big)$$
are not continuous, then the standard existence and uniqueness
theorems, see Theorems 1, 2, could not be applied. But, instead of
the ordinary differential equation (\ref{ZY}),  we can consider the
integral equation:
\begin{equation}
\label{ZY1} x(t)=x_0 + \int_{t_0}^t F(s, x(s)) ds
\end{equation}
and apply Theorems 3,4 to this equation. We note that due to the
structure of $F(t,x),$ we have in fact
$$p_1(t)=p_{01} + \int_{t_0}^t F_1 (s, q(s)) ds$$
$$p_n(t)=p_{0n} + \int_{t_0}^t F_n (s, q(s)) ds$$
$$q_1(t)=q_{01} + \frac{1}{m} \int_{t_0}^t p_1 (s, q(s)) ds$$
$$q_n(t)=q_{0n} + \frac{1}{m} \int_{t_0}^t p_n (s) ds.$$
By Propositions 1,2, $p_j(t)$ are continuous functions. Therefore
integrals $\int_{t_0}^tp_j (s) ds$ are continuous differentiable
functions. Thus under conditions of Theorem 3 or Theorem 4 we obtain
the following price dynamics:

\medskip

{\it Price trajectories are of the class $C^1$ (so
$\frac{dq}{dt}(t)$ exists and continuous), but price velocity
$$v(t)=\frac{p(t)}{m}$$ is in general non-differentiable.}

\subsection{The problem of quadratic variation}

The quadratic variation of a function $u$ on an interval $[0, T]$ is
defined as
$$
\langle u\rangle (T) = \lim_{\Vert P \Vert \to
0}\sum_{k=0}^{n-1}(u(t_{k+1})-u(t_k))^ 2,
$$
where $$P=\{0=t_0 < t_1   <...< t_{n}=T\}$$ is a partition of $[0, T]$
and $$\Vert P \Vert= \max_k \{(t_{k+1}- t_k)\}.$$  We recall the well
known result:

\medskip

{\bf Theorem} {\it If $u$ is differentiable, then $\langle f\rangle
(T) = 0.$}

\medskip

Therefore, for any smooth Bohmian trajectory its quadratic variation
is equal to zero. On the other hand, it is well known that real
price trajectories have nonzero quadratic variation, \cite{MAN},
\cite{Sh}. This is a strong objection for consideration of smooth
Bohmian price-trajectories.

In the previous section there were derived existence theorems which
provide nonsmooth trajectories. One may hope that solutions given by
those theorems would have nonzero quadratic variation. But this is
not the case.

\medskip

{\bf Theorem.} {\it Assume that,
$$
   x(t)=x_0+\int_0^t F(s,x(s))ds
$$
where $F$ is bounded, i.e., $|F(t,x)|\leq K,$ and measurable. Then
the quadratic variation $\langle F \rangle (t) = 0.$}

\medskip

{\bf Proof} We have: $$
|x(t_k)-x(t_{k-1})|^2=|\int_{t_{k-1}}^{t_k}F(s,x(s))ds|
                       \leq K^2(t_k-t_{k-1})^2.$$

Hence, with a partition of $[0,t]$, say, $0=t_0<t_1<...<t_n=t,$ we
get
$$  \sum_{k=1}^n |x(t_k)-x(t_{k-1})|^2\leq K^2\sum_1^n(t_k-t_{k-1})^2
$$
$$
\leq K^2 \max_{k:1\leq k\leq n}(t_k-t_{k-1}) \sum_1^n(t_k-t_{k-1})
 =K^2 max_{k:1\leq k\leq n}(t_k-t_{k-1}),
$$
which converges to zero as the partition gets finer, i.e. the
quadratic variation of $t\mapsto x(t)$ is zero.

\medskip

Thus the objection related to the nonzero quadratic variation is
essentially stronger than the smoothness objection. One possibility
to escape this problem is to consider unbounded quantum potentials
or even potentials which are given by distributions.

\subsection{Singular potentials and forces}

We present some examples of discontinuous quantum forces $g$
(induced by discontinuous quantum potential $U$).

\subsection{Example of singularity}

Let us consider  the wave function
$$
\psi(x)=c (x+1)^2 e^{-x^2/2}dx,
$$
where $c$ is the normalization constant providing
$$
\int_{-\infty}^{+\infty}|\psi(x)|^2 dx=1.
$$
Here $\psi(x) \equiv R(x)=|\psi (x)|.$ We have: $$ R^\prime (x)=c
[2(x+1) - x(x+1)^2] e^{-\frac{x^2}{2}}= -c(x^3 + 2x^2 -x - 2)
e^{-\frac{x^2}{2}},$$ and $$R^{\prime\prime} (x)=c
(x^4+ 2x^3 -4 x^2 -6x +1)e^{-\frac{x^2}{2}}.$$ Hence
$$
U(x)=-\frac{R^{\prime\prime} (x)}{R(x)}= \frac{x^4+ 2x^3 -4 x^2 -6x +1}{(x+1)^2}.
$$
Thus potential has singularity at the point $x=-1.$

In this example a singularity in the quantum potential $U(t, x)$ is
a consequence of division by the amplitude of the wave function
$R(t, x).$ If $|\psi (t, x_0)|=0,$ then there can appear a
singularity at the point $x_0.$

\subsection{General scheme to produce singular quantum potential for
an arbitrary Hamiltonian}

 Let $\hat H$ be a self-adjoint operator, $\hat H \geq 0,$ in $L_2
({\bf R}^n)$ (Hamiltonian -- an operator representing the financial
energy). Let us consider the corresponding Schr\"odinger equation
$$
\frac{\partial \psi}{\partial t}= \hat H \psi,$$
$$ \psi
(0)=\psi_0,$$ in $L_2 ({\bf R}^n).$ Then its solution has the form:
$$
u_t(\psi_0)=e^{\frac{-it \hat H}{h}} \psi_0.
$$
If the operator $\hat H$ is continuous, then its exponent is defined
with aid of the usual exponential power series:
$$
e^{\frac{-it \hat H}{h}}= \sum_{n=0}^\infty \Big(\frac{-it \hat
H}{h}\Big)^n/n!= \sum_{n=0}^\infty \Big(\frac{-it}{h}\Big)^n/n! \;
\hat H^n.
$$
If the operator $\hat H$ is not continuous, then this exponent can
be defined by using the {\it spectral theorem} for self-adjoint
operators.

We recall that, for any $t \geq 0,$ the map
$$u_t: L_2 ({\bf R}^n) \to L_2 ({\bf R}^n)$$ is a unitary operator:

(a) it is one-to-one;

(b) it maps $L_2({\bf R}^n)$ onto $L_2({\bf R}^n)$

(c) it preserves the scalar product:
$$ (u_t \psi, u_t \phi)=(\psi, \phi), \; \; \psi, \phi \in L_2.$$

We pay attention to the (b). By (b), for any $\phi \in L_2 ({\bf
R}^n),$ we can find a $\psi_0 \in L_2 ({\bf R}^n)$ such that
$$
\phi=u_t(\psi_0).
$$
It is sufficient to choose
$$
\psi_0=u_t^{-1} (\phi)
$$
(any unitary operator is invertible). Thus,
$$\psi(t)=u_t(\psi_0)=\phi.$$ In general a function $\phi \in L_2 ({\bf R}^n)$ is not
a smooth or even continuous function! Therefore in the case under
consideration (so we created the wave function $\psi$ such that
$\psi(t)=\phi,$ where $\phi$ was an arbitrary chosen square
integrable function),
$$
U(t, x)=- \frac{|\psi(t, x)|^{\prime\prime}}{|\psi(t,
x)|}=-\frac{|\phi(x)|^{\prime\prime}}{\phi(x)}
$$
is in general a generalized function (distribution)! For example,
let us choose
 \[ \phi(x)= \left\{ \begin{array}{ll}
\frac{1}{2b}, -b \leq x \leq b\\
0,  x \not \in [-b, b] \end{array} \right.
 \]
Here $R(t, x)=|\phi(x)|=\phi(x)$ and  $$ R^\prime (t,
x)=\frac{\delta (x + b)-\delta(x-b)}{2b},$$ $$R^{\prime\prime}(t,
x)= \frac{\delta^\prime (x + b)-\delta^\prime (x-b)}{2b}.$$

\bigskip

{\bf Conclusion.} {\it{In general, the quantum potential $U(t, x)$
is a generalized function (distribution). Therefore the price (as
well as price change) trajectory is a generalized function
(distribution) of the time variable $t$. Moreover, since the
dynamical equation is nonlinear, one cannot guarantee even the
existence of a solution.}}

\section{Classical and quantum financial randomness}

 By considering singular quantum potentials we can model the Bohmian
 price dynamics {\it with  trajectories having nonzero quadratic
 variation.} The main problem is that there are no existence theorems
 for such forces. Derivation of such theorems is an interesting
 mathematical problem, but it is completely outside of the author's
 expertise.

 Another possibility to obtain a more realistic quantum-like model
 for the financial  market is to consider additional  stochastic terms in the
 Newton equation for the price dynamics.

\subsection{Randomness from initial conditions}

Let us consider the financial Newton equation (\ref{LM}) with random
initial conditions:
\begin{equation}
\label{NE} \frac{md^2 q(t, \omega)}{dt^2}= f(t, q(t, \omega)) + g(t,
q(t, \omega)),
\end{equation}
\begin{equation}
\label{NE1} q(0, \omega)=q_0(\omega), \; \; \dot{q}(0,
\omega)=\dot{q}_0(\omega),
\end{equation}
where $q_0 (\omega)$ and $\dot{q}_0(\omega)$ are two random
variables giving the initial distribution of prices and price
changes, respectively. This is the Cauchy problem for ordinary
differential equation depending on a parameter $\omega$. If $f$
satisfy conditions of Theorem 1,  i.e., both classical and quantum
(behavioral) financial forces $f(t, q)$ and $g(t, q)$ are continuous
and satisfy the Lipschitz condition with respect to the price
variable $q$, then, for any $\omega$, there exists the solution
$q(t, \omega)$ having the class $C^2$ with respect to the time
variable $t$. But through initial conditions the price depends on
the random parameter $\omega$ so $q(t, \omega)$ is a stochastic
process. In the same way the price change $v(t, \omega)= \dot{q} (t,
\omega)$ is also a stochastic process. These processes can be
extremely complicated (through nonlinearity of coefficients $f$ and
$g).$ In general, these are {\it non-stationary processes.} For
example, the mathematical expectation
$$<q(t)>=E q (t, \omega)$$
and dispersion ("volatility")
$$
\sigma^2(q(t))=E q^2(t, \omega) - <q(t)>^2
$$
can depend on $t$.

If at least  one of financial forces, $f(t, x)$ or $g(t, x),$ is not
continuous, then we consider the corresponding integral equations:
\begin{equation}
\label{I} p(t, \omega)=p_0 (\omega) + \int_{t_0}^t f(s, q(s,
\omega)) ds + \int_{t_0}^t g (s, q(s, \omega)) ds,
\end{equation}
\begin{equation}
\label{I1} q(t, \omega)=q_0 (\omega) + \frac{1}{m} \int_{t_0}^t p(s,
\omega) ds
\end{equation}
Under assumptions of Theorem 3 or Theorem 4, there exists the unique
stochastic process with continuous trajectories, $q (t, \omega),
p(t, \omega),$ giving the solution of the system of integral
equations (\ref{I}), (\ref{I1}) with random initial conditions.

However, trajectories still have zero quadratic variation. Therefore
this model is not satisfactory.

\subsection{Random financial mass}

There parameter $m$, "financial mass", was considered as a constant
of the model. At the real financial market $m$ depends on $t$:
$$m\equiv m(t)=(m_1 (t), \ldots, m_n (t)).$$ Here $m_j (t)$ is the
volume of emission (the number of items) of shares of $j$th
corporation. Therefore the corresponding {\it market capitalization}
is given by
$$
T_j(t)= m_j(t) q_j(t).
$$
In this way we modify the financial Newton equation (\ref{NE}):
$$m_j (t) \ddot q_j=f_j (t, q) + g_j (t, q).$$
We set $F_j (t, q)=\frac{f_j (t, q) + g_j (t, q)}{m_j (t)}.$

If these functions are continuous (e.g., $m_j (t) \geq \epsilon_j >
0$ and continuous)\footnote{ The condition $m_j (t) \geq \epsilon_j
> 0$ is very natural. To be accounted at the financial market, the
volume of emission of any share should not be negligibly small.}.
and satisfy the Lipschitz condition, then by Theorem 1 there exists
the unique $C^2$-solution. If components $F_j (t, q)$ are
discontinuous, but they satisfy conditions of Theorem 3 or 4, then
there exists the unique continuous solution of the corresponding
integral equation with time dependent financial masses. By
considering the Bohmian model of the financial market with random
initial conditions it is natural to assume that even the financial
masses $m_j (t)$ are random variables, $m_j (t, \omega).$

Thus the level of emission of $j$th share $m_j$ depends on the
classical state $\omega$ of the financial market: $m_j \equiv m_j
(t, \omega).$ In this way we obtain the simplest stochastic
modification of Bohmian dynamics:
$$\ddot q_j (t, \omega)= \frac{f_j (t, q (t, \omega)) + g_j (t, q (t,
\omega))}{ m_j (t, \omega)}$$ or in the integral version:
\begin{equation}
\label{I5} q_j (t, \omega)= q_{0j} (\omega) + \int_{t_0}^t v(s,
\omega) ds
\end{equation}
\begin{equation}
\label{I6} v_j (t, \omega)= v_{0j} (\omega) + \int_{t_0}^t [f_j (s,
q (s, \omega)) + g_j (s, q (s, \omega))]/m_j (s, \omega) ds
\end{equation}

If the financial mass can become zero at some moments of time, then
{\it the price can have nonzero quadratic variation.} However, under
such conditions {\it we do not have an existence theorem.}

\section{Bohm-Vigier  stochastic mechanics}

The quadratic variation objection motivates consideration of the
Bohm-Vigier  stochastic model, instead of the completely
deterministic  Bohmian model. We follow here \cite{BOHM1}. We recall
that in the original Bohmian model the velocity of an individual
particle is given by \begin{equation} \label{BV0} v=\frac{\nabla
S(q)}{m}.
\end{equation}
If $\psi= Re^{iS/h},$ then Schr\"odinger's equation implies that
\begin{equation}
\label{BVA} \frac{dv}{dt} = - \nabla(V+U),
\end{equation}
where $V$ and $U$ are classical and quantum potentials respectively.
In principle one can work only with the basic equation (\ref{BV0}).

 The basic assumption of  Bohm and Vigier  was that the velocity
of an individual particle is given by \begin{equation} \label{BV}
v=\frac{\nabla S(q)}{m} + \eta(t),
\end{equation} where $\eta(t)$ represents a
random contribution to the velocity of that particle which
fluctuates in a way that may be represented as a random process but
with zero average.  In Bohm-Vigier the  stochastic mechanics quantum
potential comes in  through the average velocity and not the actual
one.

We now shall apply the Bohm-Vigier model to financial market, see
also E. Haven \cite{HA}. The equation (\ref{BV}) is considered as
the basic equation for the price velocity. Thus the real price
becomes a random process (as well as in classical financial
mathematics \cite{Sh}).We can write the stochastic differential
equation, SDE, for the price:
\begin{equation}
\label{BVB1} d q(t) = \frac{\nabla S(q)}{m}dt + \eta(t)dt.
\end{equation}
To give the rigorous mathematical meaning to the stochastic
differential we assume that
\begin{equation}
\label{BVB} \eta(t)= \frac{d \xi(t)}{dt},
\end{equation}
for some stochastic process $\xi(t).$ Thus formally:
\begin{equation}
\label{BVB2} \eta(t)dt=\frac{d \xi(t)}{dt} dt= d \xi(t),
\end{equation}
and the rigorous mathematical form of the equation (\ref{BVB1}) is
\begin{equation}
\label{BVB3} d q(t) = \frac{\nabla S(q)}{m}dt + d\xi(t).
\end{equation}
The expression (\ref{BVB}) one can consider either formally or in
the sense of distribution theory (we recall that for basic
stochastic processes, e.g., the Wiener process, trajectories are not
differentiable in the ordinary sense almost every where).

Suppose, for example, that the random contribution into the price
dynamics is given by {\it white noise,} $\eta_{\rm{white\;
noise}}(t).$It can be defined as the derivative (in sense of
distribution theory) of the Wiener process:
$$
\eta_{\rm{white\; noise}}(t)= \frac{d w(t)}{dt},$$ thus:
\begin{equation}
\label{BV4} v=\frac{\nabla S(q)}{m} + \eta_{\rm{white \; noise}}(t),
\end{equation}
In this case the price dynamics is given by the SDE:
\begin{equation}
\label{BVB5} d q(t) = \frac{\nabla S(q)}{m}dt + dw(t).
\end{equation}

What is the main difference from the classical SDE-description of
the financial market? This is the presence of the pilot wave
$\psi(t,q)$, mental field of the financial market, which determines
the coefficient of drift $\frac{\nabla S(q)}{m}.$ Here $S\equiv
S_\psi.$ And the $\psi$-function is driven by a special field
equation -- Schr\"odinger's equation. The latter equation is not
determined by the SDE (\ref{BVB5}). Thus, instead of one SDE, in the
quantum-like model, we have the system of two equations:
\begin{equation}
\label{BVB6} d q(t) = \frac{\nabla S_\psi(q)}{m}dt + d\xi(t).
\end{equation}
\begin{equation}
\label{SE1Z} i \; h \frac{\partial \psi}{\partial t} (t, q)= -
\frac{h^2}{2m} \frac{\partial^2 \psi}{\partial q^2} (t, q) + V(q)
\psi(t, q).
\end{equation}

Finally we come back to the problem of the quadratic variation of
the price. In the Bohm-Vigier stochastic model (for, e.g., the white
noise fluctuations of the price velocity) is nonzero.

\section{Comparison of the Bohmian model with models with stochastic volatility}
Some authors, see, e.g., \cite{Sh} for details and references,
consider the parameters of volatility $\sigma(t)$ as representing
the market behaviors. From such a point of view our financial wave
$\psi(t, q)$ plays in the Bohmian financial model the role similar
to the role of volatility $\sigma(t)$ in the standard stochastic
financial models. We recall that dynamics of $\psi(t, q)$ is driven
by the independent equation, namely the Schr\"odinger equation, and
$\psi(t, q)$ plays the role of a parameter of the dynamical equation
for the price $q(t).$

We recall the functioning of this scheme:

\medskip

a) we find the financial wave $\psi(t, q)$  from the Schr\"odinger
equation;

b) we find the corresponding quantum financial potential  $$U(t, q)
\equiv U(t, q; \psi)$$ (it depends on $\psi$ as a parameter);

c) we put $U(t, q; \psi)$ into the financial Newton equation through
the quantum (behavioral) force $g(t, q; \psi)= - \frac{\partial U(t,
q; \psi)}{\partial  q}.$

\medskip

We remark that conventional models with stochastic volatility work
in the same way, see  \cite{Sh}. Here the price $q_t$ is a solution
of the stochastic differential equation:
\begin{equation}
\label{v1} dq_t=q_t (\mu (t, q_t, \sigma_t)dt + \sigma_t
dw_t^\epsilon,
\end{equation}
where $w_t^\epsilon$ is the Wiener process, $\sigma_t$ is the
coefficient depending on time, price and volatility. And (this is a
crucial point) volatility satisfies the following stochastic
differential equation:
\begin{equation}
\label{v2} d \Delta_t=\alpha (t, \Delta_t) dt + b(t, \Delta_t)
dw_t^\delta,
\end{equation}
where $\Delta_t= ln \sigma_t^2$ and $w_t^\delta$ is a Wiener process
which is independent of $w_t^\epsilon.$

One should first solve the equation for the volatility (\ref{v2}),
then put $\sigma_t$ into (\ref{v1}) and, finally, find the price
$q_t$.

\section{Classical and quantum contributions to financial
randomness}

As in conventional stochastic financial mathematics, see, e.g.,
\cite{MAN}, \cite{Sh}, we can interpret $\omega$ as representing a
state of financial market. The only difference is that in our model
such an $\omega$ should be related to "classical state" of the
financial market. Thus we interpret conventional randomness of the
financial market  as "classical randomness", i.e., randomness that
is not determined by expectations of trades and other behavioral
factors. Besides this "classical states" $\omega$ our model contains
also "quantum states" $\psi$ of the financial market describing
market's psychology. In fact all processes under consideration
depend not only the classical state $\omega,$ but also on the
quantum state $\psi:$
 \begin{equation}
\label{LA7Z} dv_j(t,\omega, \psi)= \frac{f_j(t,q(t,\omega,
\psi),v(t, \omega, \psi), \omega)}{m_j (t,\omega)} dt +
\frac{g_j(t,q(t, \omega, \psi),\omega, \psi)}{m_j(t,\omega)} dt
\end{equation}
$$ + \sigma_j(t,\omega) d W_j (t,\omega).$$
We remark that the quantum force depends on the $\psi$-parameter
even directly:
$$
g_j= g_j(t,q,\omega, \psi).
$$
The initial condition for the stochastic differential equation
(\ref{LA7Z}) depends only on $\omega:$
$$
q_j(0,\omega)=q_{j0}(\omega), \; v_j(0,\omega)=v_{j0}(\omega).
$$
But in general the quantum state of the financial market is given
not by the pure state $\psi,$ but by the von Neumann density
operator $\rho.$ Therefore $\psi$ in (\ref{LA7Z}) is a quantum
random parameter with the initial quantum probability distribution
given by the density operator at the initial moment:
$$
\rho(0)=\rho_0.
$$
We recall that the Schr\"odinger equation for the pure state implies
the von Neumann equation for the density operator:
\begin{equation}
\label{VN} i \dot{\rho}(t)= [\hat{H}, \rho].
\end{equation}


\begin{thebibliography}{99}

\bibitem{BA}   Bachelier L 1890 \textit{Ann. Sc. l'Ecole
Normale Superiere}  \textbf{111-17} 21

\bibitem{MAN} Mantegna R N  and Stanley H E 2000 {\it   Introduction to
econophysics} (Cambridge: Cambridge Univ. Press)

\bibitem{Sh} Shiryaev A N  1999 {\it Essentials of Stochastic Finance: Facts, Models,
Theory} (Singapore: World Scientific Publishing Company)

\bibitem{SM} Samuelson P A 1965  \textit{Inductrial Management Rev.} \textbf{6}  41

\bibitem{Fam} Fama E F 1970  \textit{J. Finance} \textbf{25} 383


\bibitem{Bar} Barnett W A  and Serletis A  2000 {\it Martingales, nonlinearity,
and chaos} \textit{J. Economic Dynamics and Control} \textbf{24} 703

\bibitem{Benhab} Benhabib J 1992 {\it Cycles and Chaos in Economic Equilibrium}
(Princeton: Princeton University Press)

\bibitem{Gr} Granger C W J 1994 Is chaotic theory relevant for economics?
A review essay,  \textit{J. of International and Comparative Economics} \textbf{3}
139-145

\bibitem{Arthur} Arthur W B, Holland J H, LeBaron B, Palmer R,
and Tayler P, 1997, Asset pricing under endogenous expectations in
an artificial stock market. in W. A. Arthur, D. Lane, and S. N.
Durlauf, eds., {\it The economy as evolving, complex system-2}
(Redwood City, CA: Addison-Wesley).

\bibitem{Br} Brock W A  and Sayers C  1988 Is business cycle characterized by
deterministic chaos? \textit{Journal of Monetary Economics}
\textbf{22} 71-90

\bibitem{Cam} Campbell J Y  Lo A W  and MacKinlay A C 1997 {\it The
econometrics of financial markets} (Princeton: Princeton University
Press, Princeton)

\bibitem{BDe} DeCoster G P and D W Mitchell 1991  \textit{J. of Business and Economic Statistics} \textbf{9}
455-462

\bibitem{Hs} Hsieh D A 1991 \textit{J. of Finance} \textbf{46} 1839


\bibitem{S}  Soros J  1987 {\it The alchemy of finance. Reading of mind of the market} (New-York: J.
Wiley and Sons, Inc.)

\bibitem{TR} Tversky A and Simonson I 1993 Context-dependent preferences
\textit{Management Sc.} \textbf{39} 85-117


\bibitem{AR} Aerts D and Aerts S 1995   \textit{Foundations of
Science} \textbf{1} 1

\bibitem{AC} Accardi L 1997 {\it Urne e Camaleoni: Dialogo sulla realta, le leggi
del caso e la teoria quantistica} (Rome: Il Saggiatore)

\bibitem{KHR} Khrennikov  A Yu  2004 {\it Information dynamics in cognitive,
psychological and anomalous phenomena} (Dordreht: Kluwer)

\bibitem{BOHM1} Bohm D and Hiley B 1993 The undivided universe: an ontological
interpretation of quantum mechanics (London: Routledge and Kegan
Paul)

\bibitem{HILEY} Hiley B and  Pylkk\"anen P  1997  Active information and cognitive
science -- A reply to Kiesepp\"a, {\it  Brain, mind and physics},
eds P Pylkk\"anen, P  Pylkk\"o and A Hautam\"aki (Amsterdam: IOS
Press) p 123

\bibitem{BOHM} Bohm D  1951 {\it Quantum theory} (Englewood
Cliffs, New-Jersey: Prentice-Hall)

\bibitem{HOL} Holland P 1993 {\it The quantum theory of motion} (Cambridge: Cambridge Univ.
Press)

\bibitem{HILEY1} Hiley B 2001 From the Heisenberg picture to Bohm: a
new perspective on active information and its relation to Shannon
information {\it Quantum Theory: Reconsideration of Foundations}
ser. Math. Modelling vol. 10, ed A Yu Khrennikov  (V\"axj\"o:
V\"axj\"o University Press) p 234

\bibitem{H} Heisenberg W 1930  {\it Physical principles of quantum theory}
(Chicago: Chicago Univ. Press)

\bibitem{D} Dirac P A M 1995  {\it The Principles of Quantum Mechanics} (Oxford: Claredon
Press)

\bibitem{HA1} Haven E 2002 A discussion on embedding the Black-Scholes option pricing 
model in a quantum physics setting  \textit{Physica} A
\textbf{304} 507-524 

\bibitem{HA2} Haven E 2003 
A Black-Scholes Schrodinger option price: 'bit' versus 'qubit' 
\textit{Physica} \textbf{324}  201-206

\bibitem{HAA3} Haven E 2004 The wave-equivalent of the Black-Scholes option price: 
an interpretation \textit{Physica} A \textbf{344}  142-145

\bibitem{HA3} Haven E 2004 An 'h-Brownian motion' and the existence of stochastic option prices 
\textit{Physica} A \textbf{344} 152-155

 	 
\bibitem{SS} Segal W and  Segal I E 1998 The Black-Scholes pricing formula in the quantum context 
\textit{Proc. Nat. Acad. Sc. USA} \textbf{95} 4072

\bibitem{HAA2} Haven E 2005 Pilot-wave theory and financial option pricing. 
\textit{International J. of Theoretical Physics} \textbf{44}  1957-1962

\bibitem{HA} Haven E 2006 Bohmian mechanics in a macroscopic quantum system  {\it
Foundations of Probability and Physics-3} vol. 810 (Melville, New York: AIP) p 330


\bibitem{PPP} Piotrowski E W  Sladkowski J 2001  Quantum-like approach to financial
risk: quantum anthropic principle {\it Preprint} quant-ph/0110046

\bibitem{PP2} Piotrowski E W, Sladkowski J 2001
Quantum-like approach to financial risk: Quantum anthropic principle 
\textit{ACTA PHYSICA POLONICA} B \textbf{32} 3873-3879 

\bibitem{PP} Piotrowski E W, Sladkowski J
Quantum market games  \textit{Physica} A \textbf{312}  208-216

\bibitem{PP1} Piotrowski E W, Sladkowski J, Syska J 2003
Interference of quantum market strategies 
\textit{Physica} A \textbf{318} 516-528

\bibitem{PPP0C} Piotrowski E W, Sladkowski J 2003
Quantum English auctions  \textit{Physica} A \textbf{318}  505-515

\bibitem{PPP0A} Piotrowski E W 2003
Fixed point theorem for simple quantum strategies in quantum market games 
\textit{Physica} A \textbf{324} 196-200

\bibitem{PPP0B} Piotrowski E W, Sladkowski J 2003
Trading by quantum rules: Quantum anthropic principle 
\textit{International J. of Theoretical Physics} \textbf{42} 1101-1106

\bibitem{PPP0}  Piotrowski E W, Sladkowski J 2004
Quantum games in finance 
\textit{QUANTITATIVE FINANCE}  \textbf{4}  C61-C67 

\bibitem{PPP1} Piotrowski E W, Sladkowski J 2005
Quantum diffusion of prices and profits 
\textit{Physica} A \textbf{345}  185-195

\bibitem{PPP2} 	 Piotrowski E W, Schroeder M, Zambrzycka A 2006
Quantum extension of European option pricing based on the Ornstein-Uhlenbeck process 
\textit{Physica} A  \textbf{368} 176-182

\bibitem{DAN} Danilov V I and Lambert-Mogiliansky A 2006 Non-classical expected utility theory
{\it Preprint Paris-Jourdan Sc. Economiques}

\bibitem{DAN1} Danilov V I and Lambert-Mogiliansky A 2006 Non-classical measurement theory: a 
framework for behavioral sciences arXiv:physics/0604051

\bibitem{Man} Mandelbrot B B 1963
\textit{J. Business} \textbf{36}  394


\bibitem{KOLF} Kolmogorov A N  Fomin S V 1975 {\it Introductory Real Analysis} (New York: Dover
Publications)

\end{thebibliography}
\end{document}